\documentclass[11pt,a4paper]{article}
\pdfoutput=1

\usepackage{amsmath,amssymb}
\usepackage{epsfig,graphicx}
\usepackage{subfigure}
\usepackage{graphicx}
\usepackage{soul}
\usepackage{hyperref}
\usepackage{rotating}
\usepackage{cancel}
\usepackage{bm}
\usepackage{color}
\usepackage[font={small}]{caption}
\usepackage{comment}
\usepackage{cite}
\usepackage{psfrag}

\newcommand{\field}{\phi}

\renewcommand\[{\left[}

\newcommand{\exclude}[1]{}

\def\beq{\begin{equation}}
\def\eeq{\end{equation}}

\begin{document}
\numberwithin{equation}{section}
\title{{\huge{\bf{Gravitational waves from the fragmentation of axion-like particle dark matter
}}}
\vspace{2.5cm} 
\Large{\textbf{
\vspace{0.5cm}}}}
\author{Aleksandr Chatrchyan\thanks{chatrchyan@thphys.uni-heidelberg.de}~~and Joerg Jaeckel\\[2ex]
\small{\em Institut f\"ur theoretische Physik, Universit\"at Heidelberg,} \\
\small{\em Philosophenweg 16, 69120 Heidelberg, Germany}\\[0.5ex]
}

\date{}
\maketitle

\begin{abstract}
\noindent
The misalignment mechanism allows for the efficient, and usually very cold, production of light scalar bosons, such as axion-like particles (ALPs), making them an appealing dark matter candidate. However, in certain cases, such as in the presence of a monodromy, the self-interactions of ALPs can be sufficiently strong such that the homogeneous field fragments soon after the onset of oscillations. The resulting large inhomogeneities can lead to the production of gravitational waves (GWs).
We investigate the nonlinear dynamics of fragmentation, as well as of the subsequent turbulent regime, and calculate the stochastic GW background that is produced from this process. The GW background can be enhanced if the time evolution features an extended intermediate phase of ultrarelativistic dynamics due to a small mass at the bottom of the potential. Yet, this enhancement is limited by the requirement that the dark matter remains sufficiently cold. In some cases the resulting GWs may be within reach of future GW detectors, allowing a complementary probe of this type of dark matter.
\end{abstract}

\newpage
\label{sec:intro}
\section{Introduction}

While there is no lack of evidence for the existence of dark matter (DM) in the universe, its nature remains a mystery. Axion-like particles (ALPs) and similar light bosons, which can have masses as small as $10^{-22}\rm{eV}$, are a particularly appealing class of potential DM particles.
In field theory ALPs are typically pseudo-Goldstone bosons such as the name giving QCD axion~\cite{Peccei:1977hh, Weinberg:1977ma, Wilczek:1977pj}. But string theory, too, often features ALPs~\cite{Svrcek:2006yi, Arvanitaki:2009fg, Acharya:2010zx, Cicoli:2012sz}. Perhaps most importantly ALPs are natural DM candidates because they can be produced in the early universe via the vacuum misalignment mechanism~\cite{Preskill:1982cy, Abbott:1982af, Dine:1982ah, Arias:2012az}, in the form of a coherent field oscillating around the minimum of its potential, mimicking the behavior of pressureless matter. 

When it comes to interactions the focus is usually on those used for detection\footnote{Of course, in addition gravity allows the formation of galactic structures in the matter-dominated universe, see, e.g.~\cite{Hu:2000ke,Marsh:2015xka}.}. Yet, self-interactions can be relevant for the dynamics in the early universe, soon after the onset of oscillations when the field value is still large. For standard ALPs, whose potential enjoys a discrete shift symmetry, the effect is nevertheless usually small,\footnote{In principle also ALP--photon interactions could cause an instability that would result in a rapid decay of the ALP DM. However, in the phenomenologically relevant regime this does not happen due to a combination of the expansion of the Universe and the nonvanishing plasma mass~\cite{Alonso-Alvarez:2019ssa} (see~\cite{Abbott:1982af,Preskill:1982cy} for the case of axions).} unless the initial misalignment field value is tuned with high precision to the top of the potential~\cite{Abbott:1982af, Preskill:1982cy, Arvanitaki:2019rax}.\footnote{See~\cite{Co:2018mho, Takahashi:2019pqf} for a way to dynamically generate suitable initial conditions with the field close to the extremum of the potential.} The situation is different for ALPs exhibiting a monodromy~\cite{Jaeckel:2016qjp, Berges:2019dgr}, a phenomenon that can arise for string axions and consists in an explicit breaking of the discrete shift symmetry. Monodromy can be induced by the coupling to background fluxes~\cite{Marchesano:2014mla, Blumenhagen:2014gta, Hebecker:2014eua, McAllister:2014mpa} or branes~\cite{McAllister:2008hb, Silverstein:2008sg} wrapping a $p$-cycle that gives rise to the axion. In the presence of a monodromy the field can oscillate over multiple fundamental periods of the potential and self-interactions lead to a resonant amplification of fluctuations, as explained in~\cite{Jaeckel:2016qjp}. This generically triggers the fragmentation of the coherent field. 

We studied the nonperturbative dynamics of this process in an earlier work~\cite{Berges:2019dgr}. This confirmed the viability of such ALPs as DM for a wide range of masses. Importantly, the fragmentation significantly changes the distribution of the DM, leading to strong over-densities at small scales. While these may be difficult to detect in astrophysical and cosmological observations, they are potentially detectable in direct detection experiments looking for ALP DM, such as, e.g.~\cite{Sikivie:1983ip, Horns:2012jf, Budker:2013hfa, Jaeckel:2013sqa, Shokair:2014rna, Graham:2015ifn, Kahn:2016aff, Petrakou:2017epq, Majorovits:2017ppy, McAllister:2017lkb, Gatti:2018ojx, Melcon:2018dba, Droster:2019fur, Alesini:2019ajt,Carney:2019cio}. Fragmentation may also play an important role in the relaxion mechanism to generate the electroweak hierarchy~\cite{Graham:2015cka}, which involves similar scalar potentials. This was analyzed recently in~\cite{Fonseca:2019lmc, Fonseca:2019ypl}, where it was found that fragmentation can lead to an efficient stopping of the field.

In the following we focus on a different signature of the above mentioned dynamics caused by the self-interactions of monodromy ALP DM. The process of fragmentation and the resulting formation of strong density fluctuations accumulates a significant fraction of the energy into an anisotropic form. This can source a stochastic gravitational wave (GW) background. Such backgrounds have been discussed, e.g., in~\cite{GarciaBellido:2007af, Dufaux:2007pt, Kusenko:2008zm, Hebecker:2016vbl, Kitajima:2018zco, Lozanov:2019ylm, Sang:2019ndv}, mostly in the context of post-inflationary preheating. In particular~\cite{Hebecker:2016vbl} considered a dynamical phase decomposition after axion monodromy inflation. For ALP DM a similar mechanism for efficient GW production is provided if the ALPs have a significant coupling to the dark photon sector. The GW background produced from the decay of ALPs into dark photons was calculated in~\cite{Machado:2018nqk, Machado:2019xuc}. There, however, the authors relied on a linearized analysis. In our case, with only a single self-interacting ALP field, the nonlinear aspects play a crucial role for the fragmentation dynamics of monodromy ALPs.

We estimate the size of the GW signal analytically but also perform numerical calculations of the spectrum. For the latter we use a modified version of the ``HLATTICE'' code~\cite{Huang:2011gf}, which has been originally designed for the GW production during preheating after inflation. 

For the simplest setups, we find strong bounds on the signal strength arising from the requirement of not overproducing the DM, consistent with common expectations~\cite{Bertone:2019irm} (see also~\cite{Kitajima:2018zco}). Unfortunately this makes the detection of such a signal difficult.

Remarkably, however, an extended intermediate phase of relativistic dynamics after the fragmentation allows the production of a stronger GW signal from fragmentation and, at the same time, matching the correct final abundance of DM. Such a long relativistic phase can be achieved, if the effective mass of the particles decreases with time. In that case the evolution and the consequent increase in the fluctuations starts early in the history of the Universe. As the GW signal is roughly proportional to the square of the density this allows for a more efficient production of GWs. At the same time, during the enlarged phase of relativistic evolution the density decreases faster, still allowing it to match today's energy density. 
For monodromy ALPs such a phase can be induced by ensuring a small mass near the bottom of the potential, made possible, e.g. by higher-order periodic terms in the potential. In particular, we find that such a GW background can possibly be explored with future experiments, including pulsar timing arrays~\cite{Smits:2008cf,Moore:2014lga} as well as high-sensitivity space-based detectors~\cite{Crowder:2005nr,Audley:2017drz}, offering a promising probe of the properties of DM. 

Potentials that allow for a suitably small mass often also feature repulsive self-interactions, which is somewhat unusual for ALPs. A noteworthy qualitative change resulting from this is that, after fragmentation, a sizeable condensate re-forms\footnote{The process of condensation in the case of repulsive interactions has been studied in~\cite{Berges:2012us, Berges:2014xea, Berges:2017ldx}}.  We therefore have a combination of a quite homogeneous and a fragmented part of the ALP DM density.

Technically, the case with large mass hierarchies typically also features relatively long periods of nontrivial evolution. In an expanding universe this requires bridging a large hierarchy of scales which is difficult in a purely numerical lattice calculation. To overcome this issue we combine the numerical analysis based on lattice field theory simulations, as in~\cite{Berges:2019dgr}, with a simplified kinetic description at late times, in order to study the extended far-from-equilibrium dynamics of ALPs. The second approach becomes available due to the dramatic reduction in the complexity of the dynamics far from equilibrium~\cite{Berges:2008sr}, which is reflected in the emergence of turbulent cascades in momentum space with universal scaling behavior and self-similar time evolution~\cite{Micha:2004bv, Berges:2014xea, Orioli:2015dxa,  Berges:2017ldx}. This allows us to estimate the duration of the relativistic phase, as well as the typical velocities of ALPs at matter-radiation equality, without performing extensive lattice simulations of the full dynamics. 

Let us briefly outline the main steps of the paper. In Sec.~\ref{sec:outline} we recall the main stages of the dynamics and the mechanism for the production of GWs. In Sec.~\ref{sec:analytics} we perform analytical estimates for the GW signal and describe the bounds from total DM abundance. We describe our numerical analysis in Sec.~\ref{sec:numsim}. In sections~\ref{sec:simple} and \ref{sec:extended} we discuss in detail the dynamics for the simple scenario and the one with an extended relativistic phase, respectively. Sec.~\ref{sec:conc} is used to give some brief conclusions. In terms of conventions we take $\hbar=c=1$ and a Friedmann-Robertson-Walker (FRW) metric $ds^2=dt^2-a^2(t)d\mathbf{x}^2$ ($a(t)$ is the scale factor) which we also use in conformal time $ds^{2}=a^{2}(t)(d\tau^2-d\mathbf{x}^2)$.

\section{Outline of the dynamics}
\label{sec:outline}

In this section we review the production of ALP DM. We focus on the regime when the self-interactions are important and lead to nonperturbative dynamics, including the fragmentation of the initially homogeneous ALP field. We also review the linearized theory of gravity, which is used to calculate the stochastic GW background from this process.

\subsection{Misalignment production of ALPs and fragmentation}

In this work we consider the scenario of the misalignment mechanism, in which the ALP field is present during inflation (see, e.g.~\cite{Arias:2012az, Marsh:2015xka, Berges:2019dgr}). Initially the field takes some value, which becomes homogeneous throughout the observable universe during inflation. We denote this value by $\phi_1$. The field is effectively frozen and starts to perform coherent oscillations once the Hubble parameter becomes comparable to the curvature of the potential~\cite{Marsh:2015xka}.

The classical field equations of motion in an expanding FRW spacetime read
\beq
\label{eq:FRWeq}
\ddot \field + 3 H \dot \field - \frac{\Delta \field}{a^2}  +\frac{\delta U}{\delta \phi} =0,
\eeq
where $a(t)$ is the scale factor, $H=\dot a/a$ is the Hubble parameter and $U(\phi)$ is the potential of the ALPs.

In the presence of a monodromy~\cite{McAllister:2008hb, Silverstein:2008sg, Marchesano:2014mla, Blumenhagen:2014gta, Hebecker:2014eua, McAllister:2014mpa} the potential consists of a periodic term and one that breaks the periodicity. The first term can be typically parameterized as $U( \phi )  = \Lambda^4 [1-\cos ( \phi / f) ]$, where $f$ is the so-called decay constant and $\Lambda$ is the scale of the nonperturbative effects that generate the ALP potential. The second term can be some monomial. For simplicity, and as in~\cite{Jaeckel:2016qjp, Hebecker:2016vbl}, we consider a quadratic monomial breaking the periodicity, of the form
\beq
\label{eq:monopot}
U(\phi)=\frac{1}{2}m^2\phi^2 + \Lambda^4 \Bigl[ 1-\cos \left( \frac{\phi}{f}+\delta \right) \Bigr].
\eeq
In principle the phase $\delta$ can take nontrivial values. However, for most of the discussion we will take $\delta=0$ for simplicity.
We will mention explicitly when considering a nonvanishing phase $\delta$.

The dynamics in such a potential was considered in~\cite{Jaeckel:2016qjp, Berges:2019dgr}. Importantly, the breaking of periodicity allows to have large field displacements over multiple fundamental periods, \mbox{$\phi_1/f \gg 1$}. 

In the presence of oscillations the self-interactions of ALPs lead to parametric instabilities~\cite{Berges:2002cz}. These result in an amplification of the fluctuations of the field, which are $\delta \phi \sim H_I$ at the onset of oscillations, where $H_I$ is the inflationary Hubble parameter. For sufficiently large misalignment angles $\phi_1/f$ these fluctuations get amplified to $\delta \phi \sim f$, making the dynamics fully nonperturbative\footnote{Note that $\phi_{1}\gg H_I$ must hold if ALPs are present during inflation in order to avoid isocurvature fluctuations~\cite{Visinelli:2009zm,Arias:2012az,Alvarez:2017kar}.}.  
The transfer of energy from the coherently oscillating field to the fluctuations continues and eventually leads to the complete fragmentation of the ALP field. Due to the large field values/occupation numbers, this dynamics can be well described in terms of classical(-statistical) field theory, as it was done in~\cite{Berges:2019dgr}.

Directly after fragmentation ALPs are usually relativistic, such that their energy density dilutes efficiently, as $\sim a^{-4}$. The dynamics at this stage is strongly nonlinear, involving turbulent cascades in momentum space~\cite{Micha:2004bv, Berges:2014xea, Orioli:2015dxa,  Berges:2017ldx}. As ALPs dilute due to Hubble expansion these nonlinear effects gradually become unimportant. ALPs also cool down with expansion and become nonrelativistic once their characteristic physical momenta drop below their mass. After this they can successfully participate in structure formation.

\begin{figure}[!t]
	\centering
	\includegraphics[width=0.99\textwidth]{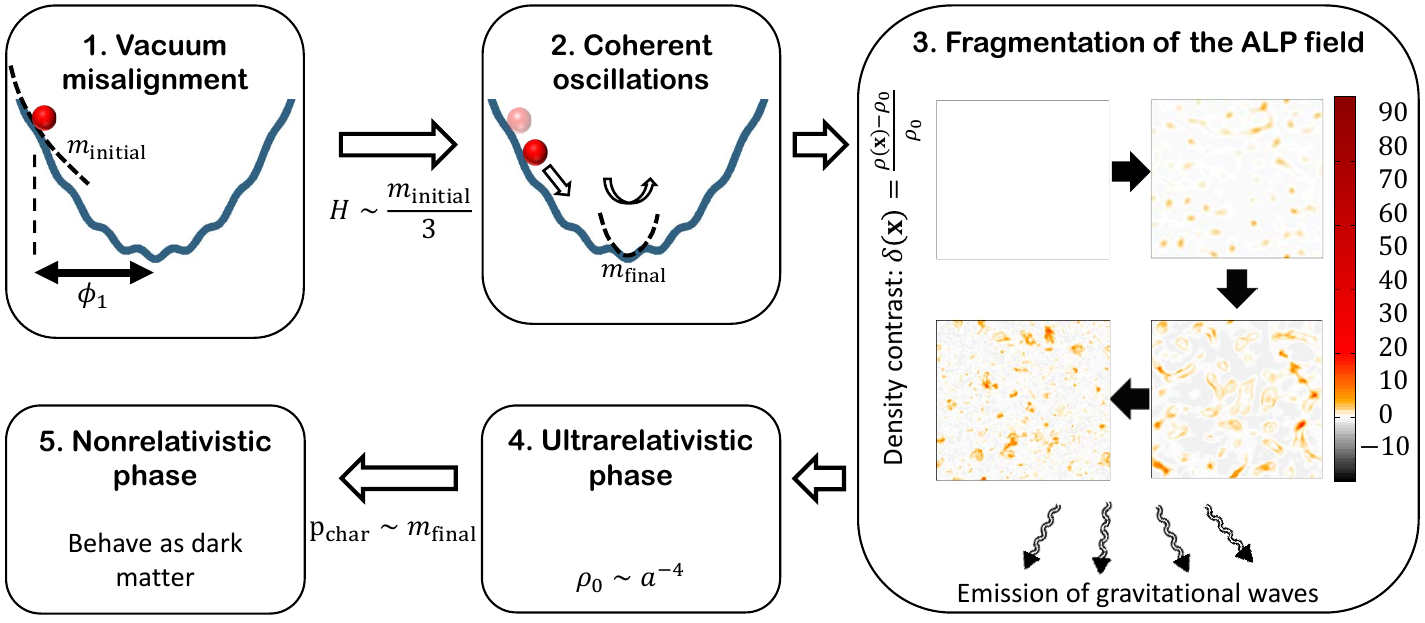}
	\caption{The main stages of the misalignment production of ALPs, involving nonperturbative dynamics. Initially the field is frozen (1). It starts to oscillate (2) after the Hubble parameter $H$ drops below the curvature of the potential $m_{\rm initial}$. This is followed by the fragmentation of the ALP field (3), a process that produces a stochastic GW background and is illustrated by plotting several snapshots of the energy density contrast $\delta(\mathbf{x}) = (\rho(\mathbf{x})-\rho_0)/\rho_0$ along a 2D slice. After the fragmentation ALPs are relativistic (4) and become nonrelativistic (5) after their characteristic physical momenta $\rm p_{char}$ drop below the curvature at the bottom of the potential $m_{\rm final}$.}
	\label{stages}
\end{figure}

This sequence of stages of the dynamics (see also~\cite{Berges:2019dgr} for details) is illustrated in Fig.~\ref{stages}. In particular, in the third panel, which describes the fragmentation of the ALP field, several snapshots of the energy density contrast $\delta(\mathbf{x}) = (\rho(\mathbf{x})-\rho_0)/\rho_0$ along a 2D slice are shown. These are obtained from a classical lattice simulation of the nonlinear field dynamics according to (\ref{eq:FRWeq}).

In Fig.~\ref{stages}, the distinction between two different mass parameters is made explicit. One of them, which we denote as $m_{\rm initial}$, is the typical curvature of the potential near $\phi_1$ and determines the onset of coherent oscillations via the condition $H \sim m_{\rm initial}/3$. The other parameter, denoted as $m_{\rm final}$, is the curvature at the bottom of the potential and the transition to the nonrelativistic regime at late times happens when the characteristic momenta drop below this mass, $\mathrm{p}_{\rm char} \sim m_{\rm final}$. For the monodromy potential in \eqref{eq:monopot} this mass, i.e. the curvature near the bottom is equal to 
\beq
m_a^2 = m^2 + \frac{\Lambda^4}{f^2} = m^2 (1+\kappa^2), \: \: \: \: \: \: \:  \: \: \: \: \: \:\kappa^2 = \frac{\Lambda^4}{m^2 f^2}.
\eeq
Assuming values $0\lesssim\kappa^2\lesssim {\rm few}$, $m_{\rm initial}$ and $m_{\rm final}$ are of similar size. However, this is not necessary for generic monodromy potentials. For more general potentials the mass at the bottom can be different and, in particular, significantly smaller. This would lead to an extended ultrarelativistic phase and will be important in the context of GW production, as we argue in Sec.~\ref{subsec:parametric} and demonstrate in Sec.~\ref{sec:extended}.
One option to achieve such a small mass is to flip the sign in front of $\kappa^2$ in the above equation. This corresponds to setting $\delta=\pi$ in Eq.~\eqref{eq:monopot}. Then taking $\kappa^2$ close to $1$ but still smaller than $1$, the global minimum remains at $\phi=0$ but the mass in its vicinity is very small. Alternatively, in more general situations with multiple periodic contributions to the potential this can also be achieved. In Sec.~\ref{sec:extended} we consider both cases but with a focus on the more general second case.

\subsection{Today's abundance of DM}

Let us at this point introduce some notation that will be used later. We denote by $a_{\rm osc}$ the value of the scale factor at the moment when the Hubble parameter is equal to $H_{\textrm{osc}} = m_a/3$. This is approximately the scale factor after which the field enters the oscillatory regime. 

In addition we also introduce a rescaled dimensionless comoving momentum 
\begin{equation}
\label{eq:etadef}
\eta = \frac{\mathrm{p}}{m_a a_{\rm osc}},
\end{equation}
that will be frequently used in this work. 

The ALP energy densities today and at $a_{\rm osc}$ are related to each other via
\beq
\label{eq:dmdensitytd_0}
\rho_{\phi, \mathrm 0} = \rho_{\phi, \mathrm{osc}} \Bigl( \frac{a_{\mathrm{osc}}}{a_0} \Bigr)^3 \mathcal{Z}^{\mathrm{osc}}_0.
\eeq
Here the dimensionless prefactor
\beq
\label{eq:Z}
\mathcal{Z}_{a_2}^{a_1} = \exp \Big[-3\int_{a_{1}}^{a_2} w_{\phi}(\widetilde a) d\ln \widetilde a   \Bigr],
\eeq
which was also used in~\cite{Berges:2019dgr}, is determined from the ALP equation of state parameter, $w_{\phi}=\langle p_{\phi} \rangle/ \langle \rho_{\phi} \rangle$. It describes how much the energy density of ALPs is suppressed on top of the $\propto a^{-3}$ dilution of nonrelativistic matter. This can be caused by an intermediate ultrarelativistic phase, which leads to $w_{\phi}=1/3$. 

Today's ALP energy density (\ref{eq:dmdensitytd_0}) can be expressed in terms of the misalignment field value and the mass via the equation (4.16) from~\cite{Berges:2019dgr},
\beq
\label{eq:dmdensitytd}
\rho_{\phi, \mathrm 0}  = 0.17\, \frac{\mathrm{keV}}{\mathrm{cm}^3} \sqrt{\frac{m_{a}}{\mathrm{eV}}} \Bigl(\frac{\phi_1}{10^{11}\mathrm{GeV}}\Bigr)^2 \Bigl( \frac{\mathcal{Z}_{0}^{\mathrm{osc}}}{1+\kappa^2} \Bigr) \mathcal{F}(T_{\mathrm{osc}}),
\eeq
where the dimensionless factor $\mathcal{F}(T_{\rm osc}) = ( g_{s,0} / g_{s,\mathrm{osc}}  ) (  g_{\mathrm{osc}} /g_{0}  )^{3/4}$ encodes the changing number of degrees of freedom for entropy, $g_{s}(T)$, and energy, $g(T)$,\footnote{Their values can be found, e.g., in~\cite{Husdal:2016haj}.} and takes values between $1$ and $0.3$ depending on the temperature~\cite{Arias:2012az}. Inserting~\cite{Aghanim:2018eyx} the value of today's DM density $\rho_{DM}\sim 1.27 \, \mathrm{keV}/{\mathrm{cm}^3}$, one arrives at the following expression for the final relic abundance relative to the observed DM
\begin{equation}
\frac{\Omega_{\phi}}{\Omega_{\rm DM}} \simeq 13 \sqrt{ \frac{m_a}{\rm{eV}}   } \Bigl( \frac{\phi_1}{10^{12}\rm{GeV}} \Bigr)^2 \Bigl( \frac{\mathcal{Z}^{\rm{osc}}_{\rm{0}} }{1+\kappa^2} \Bigr) \mathcal{F}(T_{\rm osc}).
\end{equation}
This approximately coincides with Eq.~(4) of~\cite{AlonsoAlvarez:2019cgw} if one inserts the value of $(\mathcal{Z}^{\rm osc}_0/(1+\kappa^2))\approx 0.5$ that is observed for small misalignment angles (see Fig.~7 of \cite{Berges:2019dgr}).

\subsection{Linear metric fluctuations}

GWs correspond to transverse-traceless (TT) tensor-like perturbations of the metric tensor. In the synchronous gauge, where $g_{00}=1$ and $g_{0i}=0$, these perturbations on top of the background FRW metric have the form~\cite{GarciaBellido:2007af, Dufaux:2007pt},
\beq
g_{\mu \nu}(t, \mathbf{x}) = \Biggl( \begin{array}{cc} 1 & 0 \\ 0 & -a^2(t)\Bigl( \delta_{ij}+h_{ij}(t, \mathbf{x}) \Bigr) \end{array} \Biggr),
\eeq
with $|h_{ij}| \ll 1$. The equations of motion for $h_{ij}$ are obtained by linearizing the Einsteins field equations~\cite{carroll2019spacetime},
\beq
\label{gweq}
\ddot h_{ij}+3 H \dot h_{ij} - \frac{\Delta h_{ij}}{a^2}  =  \frac{16 \pi}{M_{\mathrm{Pl}}^2} \Pi^{TT}_{ij}.
\eeq
The TT condition, $\partial_i h_{ij}=h_{ii}=0$, leaves the metric perturbations with two independent components, which correspond to the two polarizations of GWs. $\Pi^{TT}_{ij}$ is the TT projection of the total anisotropic energy-momentum tensor which, for a scalar field is given by~\cite{GarciaBellido:2007af, Dufaux:2007pt}
\beq
\Pi_{ij}(t,\mathbf{x}) = \frac{1}{a^2} \Bigl[\partial_i \phi(t,\mathbf{x}) \partial_j \phi(t,\mathbf{x}) -\delta_{ij}\Bigr(\mathcal{L}(\phi(t,\mathbf{x}))-\langle p \rangle \Bigl) \Bigr],
\eeq
where $\mathcal{L}$ is the Lagrange density and $\langle p \rangle$ the background pressure of the universe. The components, proportional to $\delta_{ij}$ drop out after taking the TT projection,
\begin{equation}
\Pi^{TT}_{ij}(t,\mathbf{x})=\frac{1}{a^2}\left[\partial_i \phi(t,\mathbf{x}) \partial_j \phi(t,\mathbf{x})-\frac{1}{3}\delta_{ij}\left(\partial_k \phi(t,\mathbf{x}) \partial_k \phi(t,\mathbf{x}) \right)\right].
\end{equation}

In other words, GWs are sourced only by the gradients of the field, which explains why the process of fragmentation is an important source of GWs.

\section{Qualitative analysis of GW production from ALP DM}
\label{sec:analytics}
Having set up the basic equations governing the system, in this section we estimate analytically the strength and peak frequency of the GW signal produced during the nonperturbative ALP dynamics. The strength of such stochastic backgrounds is conveniently described in terms of the dimensionless parameter $\Omega_{\rm GW}$, which determines the fraction of energy density in GWs per logarithmic frequency to the total energy density of the universe,
\beq
\label{en_frac_formula}
\frac{\rho_{\mathrm{GW}}}{\rho_{\mathrm{c}}} = \int d(\ln\nu)\,\,  \Omega_{\mathrm{GW}}(\nu),
\eeq
where $\nu$ is the frequency, $\rho_c = (3/8\pi) M_{\rm Pl}^2 H^2$ is the critical density and $\rho_{\rm GW}$ is the energy density in GWs, which is given by\footnote{In these expressions summation over the indices $i$ and $j$ is implied.}~\cite{GarciaBellido:2007af}
\beq
\label{enGW}
\rho_{\mathrm{GW}}(t) = \frac{M_{\mathrm{Pl}}^2}{32 \pi} \langle  \dot h_{ij}(t,\mathbf{x})  \dot h_{ij}(t,\mathbf{x}) \rangle = \frac{M_{\mathrm{Pl}}^2}{32 \pi V} \int_{\mathbf{p}} \Bigl( |\dot h_{ij}(t,\mathbf{p})|^2  \Bigr).
\eeq

\bigskip
After describing the transformation of the frequency and the energy fraction of the GW signal to today's variables we estimate its strength from (\ref{gweq}). We then fix the energy density of ALPs based on today's DM abundance to find the signal strength from ALP DM.

\subsection{Transformation to today's observables}
\label{sec:dm_today}

Today's frequency of GWs is related to the comoving momentum $\eta$ by a simple red-shift,
\beq
\label{GW_nu_dm}
\nu = \frac{1}{2\pi}\frac{\mathrm{p}}{a_{\rm emit}}\frac{a_{\rm emit}}{a_{0}}
=\frac{1}{2\pi}\frac{{\rm p}}{a_0} = \frac{\eta m_a }{2\pi} \Bigl( \frac{a_{\rm osc}}{a_0} \Bigr).
\eeq
Here $\mathrm{p}/a_{\rm emit}$ is the physical momentum at the time when emission takes place. Then the factor $a_{\rm emit}/a_{0}$ takes care of the red-shift till today. This is then expressed using the rescaled characteristic momentum scale as defined in Eq.~\eqref{eq:etadef}.  

For the ratio of the scale factors we then use the conservation of comoving entropy,
$a_{\rm osc}/a_0  = (T_0/T_{\rm osc}) ( g_{s,0}/g_{s, \rm osc} )^{1/3}$,
where $T_0 = 2.73 \mathrm{K}$ and $T_{\rm osc} = \sqrt{H_{\rm osc} M_{\rm Pl} / (1.66 \sqrt{g_{\rm osc}} ) }$ during the epoch of radiation-domination. Recalling that $H_{\rm osc} = m_a/3$, the formula  (\ref{GW_nu_dm}) can be re-written as 
\beq
\label{nu_today}
\nu = \Bigl( \frac{\sqrt{3}\sqrt{1.66}g_{0}^{1/4} T_0 \sqrt{\mathrm{eV}} }{2 \pi \sqrt{M_{\rm Pl}} \mathrm{Hz}} \Bigr)\mathrm{Hz} \, \eta \,  \sqrt{ \frac{m_a}{\mathrm{eV}} } \mathcal{F}^{1/3}(T_{\rm osc}) = 1.56 \times 10^{-3} \mathrm{Hz} \, \eta \,  \sqrt{ \frac{m_a}{\mathrm{eV}} } \mathcal{F}^{1/3}(T_{\rm osc}).
\eeq
As can be seen the frequency is mostly determined by the ALP mass.

The energy density in GWs dilutes as $a^{-4}$, hence today's energy fraction $\Omega_{\rm GW, 0}$ can be expressed as~\cite{Machado:2018nqk},
\beq
\label{GW_omega_dm}
\Omega_{\rm GW, 0} = \frac{\rho_{0}}{\rho_{c, 0}} = \Omega_{\rm GW, emit} \Bigl( \frac{ a_{\rm{emit}} }{ a_{0} }\Bigr)^{4} \Bigl( \frac{H_{\mathrm{emit}}}{H_0} \Bigr)^{2} = \Bigl( \frac{1.66^2 \times g_{0} T_0^4}{M_{\rm Pl}^2 H_0^2}\Bigr) \Omega_{\mathrm{GW}, \rm emit} \mathcal{F}^{4/3}(T_{\rm emit}),
\eeq
where we again used the conservation of comoving entropy. Calculating the term in the brackets, the formula (\ref{GW_omega_dm}) can be re-written as
\beq
\label{om_today}
\Omega_{\mathrm{GW}, 0} = 9.39 \times 10^{-5} \Omega_{\mathrm{GW}, \rm emit}  \mathcal{F}^{4/3}(T_{\mathrm{emit}}).
\eeq
Equations (\ref{nu_today}) and (\ref{om_today}) are used in our numerical simulations to transform the variables to today's observables, as well as for our analytical estimates of GW production, described below.

\subsection{Parametric estimates for the GW signal}\label{subsec:parametric}

We now perform parametric estimates for the GW production. For simplicity, we assume that most of it takes place near some scale factor $a_{\rm emit}$. This is naturally identified with the end of the resonant amplification of fluctuations and the fragmentation of the background field (the third stage in Fig.~\ref{stages}). We denote by $\mathrm{p}_{\star} = \eta_{\star} m_a a_{\rm osc}$ the comoving momentum at which the dominant resonant production takes place, and by $\nu_{\star}$ the corresponding frequency. For our parametric estimates we assume that the GW spectrum is peaked at a momentum of the same order as $\mathrm{p}_{\star}$ and, in what follows, estimate the energy fraction at that frequency. 

From the wave equation (\ref{gweq}) it follows that~\cite{Giblin:2014gra}
\beq
\label{whatish}
| (h_{\mathrm{p}_{\star}, \rm emit})_{ij} | \sim \frac{16 \pi  }{M^2_{Pl} (\mathrm{p}_{\star} / a_{\rm emit})^2 } | (\Pi_{\mathrm{p}_{\star}, \rm emit})^{TT}_{ij} |.
\eeq
For simplicity we ignore the index structure of metric perturbations, i.e.~identify $h_{ij} \rightarrow h$, and associate the $\Pi^{TT}_{ij}$ to the energy density of the ALP field, i.e. 
\begin{equation}
\Pi^{TT}_{ij} = \alpha \rho_{\phi}, 
\end{equation}
where
$$
\alpha \lesssim 1, 
$$
roughly quantifies the fraction of the energy stored in the fluctuations.
These are of course crude approximations, but are sufficient for a parametric estimate.

The energy density $\rho_{\mathrm{GW}, \rm emit}$ in GWs at $a_{\rm emit}$ can be written as
\beq
\rho_{\mathrm{GW}, \rm emit} = \frac{M_{Pl}^2}{32\pi V}  \int  |\dot h_{\rm p}|^2\frac{4\pi}{(2\pi)^3}  {\rm p}^2 d{\rm p}   \approx \frac{M_{Pl}^2}{8 V (2\pi)^3}  \int d{(\ln\rm p)} {\rm p}^3 \Bigl( \frac{\rm p}{a_{\rm emit}}\Bigr)^2 | h_{\rm p}|^2.
\eeq
Here $V$ indicates the control volume for the Fourier transform which will drop out of the final expressions.

The GW energy fraction at the momentum $\rm p_{\star}$ is then given by
\beq
\label{GW_emit}
\Omega_{\mathrm{GW}, \rm emit}(\nu_{\star}) \approx \frac{M_{Pl}^2}{8  \rho_{c, \rm emit}V(2\pi)^3} \frac{\rm p_{\star}^5}{a^2_{\rm emit}} | h_{\rm p_{\star}, emit}|^2 \sim \frac{16^2 \pi^3}{3 M_{Pl}^4 H_{\rm emit}^2 V(2\pi)^3} \alpha^2|\rho_{\phi, \rm p_{\star}, emit}|^2 {\rm p}_{\star} a^2_{\rm emit} ,
\eeq
where in the last step we inserted (\ref{whatish}) as well as expressed the critical density at $a_{\rm emit}$ in terms of the Hubble parameter $H_{\rm emit}^2 = 8\pi\rho_{c, \rm emit}/(3M_{Pl}^2)$.

Noting that the spectra are typically peaked at the resonant momentum, we can employ Parseval's relation to re-express the Fourier components $\rho_{\phi, \mathrm{p}_{\star}, \rm emit}$ in terms of position space ALP energy density $\rho_{\phi, \rm emit}$, 
\begin{equation}
V \rho_{\phi, \rm emit}^2 = \int_\mathbf{p} |\rho_{\phi, \mathrm{p}, \rm emit}|^2 = \frac{4\pi}{(2\pi)^3} \int d (\ln{\rm p})  |\rho_{\phi, \mathrm{p}, \rm emit}|^2 {\rm p}^3 \approx \frac{4\pi}{(2\pi)^3} |\rho_{\phi, \mathrm{p}_{\star}, \rm emit}|^2  \mathrm{p}_{\star}^3 \beta.
\end{equation}
In this equation we used the typical logarithmic width of the spectrum of the fluctuations in momentum space, 
\begin{equation}
\beta = \Delta \ln{\rm p},
\end{equation}
where usually $\beta \gtrsim 1$. Inserting this into (\ref{GW_emit}) yields
\beq
\label{eq:omega1}
\Omega_{\mathrm{GW}, \rm emit}(\nu_{\star}) \sim  \frac{64 \pi^2}{3 M_{Pl}^4 H^2_{\rm emit} } \frac{\rho_{\phi, \rm emit}^2}{ (\mathrm{p}_{\star}/a_{\rm emit})^2}\frac{\alpha^2}{\beta}.
\eeq

It is convenient to re-express the energy fraction in terms of quantities at $a_{\rm osc}$. For this we note that 
\begin{equation}
\label{eq:hemit}
H_{\rm emit} = H_{\rm osc} \Bigl( \frac{a_{\rm osc}}{a_{\rm emit}}\Bigr)^2 = \frac{m_a}{3} \Bigl( \frac{a_{\rm osc}}{a_{\rm emit}}\Bigr)^2,\: \: \: \: \: \: \: \: \: \: \: \: \: \: \: \rho_{\phi, \rm emit} \approx \rho_{\phi, \rm osc}\Bigl( \frac{a_{\rm osc}}{a_{\rm emit}}\Bigr)^3 \mathcal{Z}^{\mathrm{osc}}_{\rm emit}.
\end{equation}
The factor $\mathcal{Z}^{\mathrm{osc}}_{\rm emit}$ in the second equation (and defined in Eq.~\eqref{eq:Z}) accounts for the nontrivial evolution of the ALP energy density between $a_{\rm osc}$ and $a_{\rm emit}$ compared to the standard $\propto a^{-3}$ dilution for matter. This evolution results from the strong fluctuations which correspond to particles with nonnegligible momenta. 

Inserting Eq.~\eqref{eq:hemit} into Eq.~\eqref{eq:omega1} yields
\beq
\label{Omegatd}
\Omega_{\mathrm{GW}, 0}(\nu_{\star}) \sim 0.18 \frac{\rho_{\phi, \rm osc}^2 }{M_{Pl}^4 m_a^4 \eta_{\star}^2} (\mathcal{Z}^{\rm osc}_{\rm emit})^2 \mathcal{F}^{4/3}(T_{\mathrm{emit}}) \frac{\alpha^2}{\beta}.
\eeq

Finally, inserting (\ref{eq:dmdensitytd_0}) into (\ref{Omegatd}) with the value $\rho_{\phi, 0}=1.27 \, {\mathrm{keV}}/{\mathrm{cm}^3}$ for the DM energy density today~\cite{Aghanim:2018eyx} leads to
\beq
\label{anal_fin}
\Omega_{\mathrm{GW}, 0}(\nu_{\star}) \sim 1.1 \times 10^{-32} \Bigl[ \frac{\rm eV}{m_a} \Bigr] \Bigl( \eta_{\star} \mathcal{Z}_0^{\rm emit}\mathcal{F}^{1/3}(T_{\rm osc})\Bigr)^{-2}  \frac{\alpha^2}{\beta} ,
\eeq
where we employed the conservation of comoving entropy and the relation $\mathcal{Z}^{\mathrm{osc}}_0 = \mathcal{Z}^{\mathrm{osc}}_{\rm emit} \mathcal{Z}^{\mathrm{emit}}_0$ which follows directly from the definition (\ref{eq:Z}). We also assume for the sake of simplicity that $\mathcal{F}(T_{\rm osc})\approx \mathcal{F}(T_{\rm emit})$.

Eq.~(\ref{anal_fin}) determines the peak GW signal strength for a given ALP mass. Combining this expression with Eq.~(\ref{nu_today}) for the peak frequency, $\nu_{\star}/{\mathrm{Hz}} = 1.56 \times 10^{-3}  \eta_{\star}  \sqrt{ m_a/ \mathrm{eV}} \mathcal{F}^{1/3}(T_{\rm osc})$, one arrives at a remarkably simple relation between the peak energy fraction $\Omega_{\mathrm{GW}, 0, \star} = \Omega_{\mathrm{GW}, 0}(\nu_{\star})$ and the peak frequency $\nu_{\star}$, 
\beq
\label{signal_str_form}
\Omega_{\mathrm{GW}, 0}(\nu_{\star}) \sim 2.7\times 10^{-38}   \frac{1}{(\nu_{\star}/{\rm Hz})^2} \frac{1}{ (\mathcal{Z}_0^{\rm emit})^2} \frac{\alpha^2}{\beta}.
\eeq

\subsection{Summary of the analytical results}
\label{ssec:summary}

The estimates of the signal strength in the previous subsections were very general and did not rely on details of the ALP potential $U(\phi)$. The main requirement was that the typical curvature of the potential is given by $m_a$. It was, however, assumed that the spectrum is centered close to the resonant momenta. Nonlinear effects tend to broaden the spectrum and, as we demonstrate in the next section, somewhat modify this picture. Nevertheless, (\ref{signal_str_form}) reveals the important features of the GW spectrum.

It follows from (\ref{signal_str_form}) that on top of the inverse-squared dependence on the frequency, the peak energy fraction is determined solely from the value of the suppression factor $\mathcal{Z}_0^{\rm emit}$, given by (\ref{eq:Z}). If $\mathcal{Z}_0^{\rm emit} = \mathcal{O}(1)$, which is the case for the potential of (\ref{eq:monopot}) (the values of $\mathcal{Z}_0^{\rm emit}$ can be found in~\cite{Berges:2019dgr}), this signal peak is beyond the planned sensitivities of future GW detection experiments, irregardless of the value of the ALP mass. We investigate this scenario numerically in Sec.~\ref{sec:simple}. In particular, in Fig.~\ref{spec_gw_att} the analytical estimate, denoted by the blue circle, is shown together with the numerically calculated GW spectrum for $m_a=10^{-16}\rm{eV}$ and a particular set of parameters. The (numerically extracted) final GW spectra for different values of parameters are shown in Fig.~\ref{gw1}.

This bound on the signal strength is lifted if ALPs dilute their energy more efficiently after the emission of GWs, i.e. if $\mathcal{Z}_0^{\rm emit} \ll 1$. Indeed, a value of $\mathcal{Z}_0^{\rm emit}$, sufficiently smaller than one, implies that the ALP field constituted a larger fraction of the total energy of the universe when GWs were mostly emitted and, therefore, the signal carried an accordingly larger energy fraction. One possibility how a small value of $\mathcal{Z}_0^{\rm emit}$ can be achieved is if ALPs exhibit an extended phase of relativistic dynamics after fragmentation (stage 4 in Fig.~\ref{stages}), which can be induced by the bare mass at the bottom of the potential $m_{\rm final}$ being smaller than $m_a$. We focus on this scenario in Sec.~\ref{sec:extended}

Our estimate of the signal strength is very general. It depends solely on the requirement of producing the observed amount of DM. Obviously there are additional constraints, mostly from large-scale structure formation probes. These require ALPs to be sufficiently cold/pressureless at matter-radiation equality. The nonrelativistic velocities at matter-radiation equality $\mathrm{v_{eq}}$ can be expressed in terms of the comoving momenta via 
\beq
\label{exp_for_v}
\mathrm{v_{eq}} = \frac{\rm p}{a_{\rm eq} m_{\rm final}} = 2.63 \, \eta  \sqrt{\frac{10^{-28}\mathrm{eV}}{m_{\rm final}} }  \sqrt{ \frac{m_a}{m_{\rm final}} } \frac{ \mathcal{F}^{1/3}(T_{\rm osc}) }{ \mathcal{F}^{1/3}(T_{\rm eq}) } ,
\eeq
where we again used entropy conservation and inserted $H_{\rm eq} = 2.3\times 10^{-28}{\rm eV}$.

Our analytical estimates are expected to become less accurate for larger misalignment angles $\phi_1/f$, if there is more time between the end of resonant amplification and the final fragmentation. In that case the signal peak is expected to be weaker and, instead, the signal to be broader. Moreover, these estimates do not take into account the production of GWs after fragmentation, during the turbulent dynamics of ALPs. Finally, the calculation of $\mathcal{Z}_0^{\rm emit}$ and of the typical velocities at matter-radiation equality requires a better knowledge of the ALP spectrum. This motivates the more rigorous analysis based on numerical lattice simulations, which is given in the next sections.

\section{GW spectra from numerical simulations}
\label{sec:numsim}

In this section we describe our numerical implementation, based on the classical-statistical approximation~\cite{Berges:2015kfa}.

The simulation of the dynamics of the scalar field is done similar to our previous work~\cite{Berges:2019dgr}, to which we refer for more details. Linear metric perturbations and the corresponding GW spectra are computed using the HLATTICE code~\cite{Huang:2011gf}. The code is originally designed for the simulation of post-inflationary preheating dynamics. To adjust it for the case of misalignment production of ALP DM, the following modifications were performed:
 
\begin{itemize}
\item The field evolves on a radiation-dominated background, $a \sim \sqrt{t}$, without having any impact on this background expansion rate.
\item The horizon entry of frozen modes is properly taken into account, by ``gluing'' the lattice simulation to an early linear evolution, as described in~\cite{Berges:2019dgr}. In other words, we first evolve the momentum modes of the fluctuations linearly, starting with the scale-independent power spectrum (cf. e.g.~\cite{Baumann:2009ds}),
\beq
\Delta_{\field}= \Bigl(\frac{H_I}{2\pi}\Bigr)^2,
\eeq
imprinted after inflation, and then,  prior to the onset of nonlinear effects, switch to the lattice simulation. In this way we also neglect the GW production during this early linear phase. However the latter usually only gives strongly (exponentially) suppressed contributions.
\item We always assume that ALPs constitute the total abundance of DM. For given values of the ALP masses and the misalignment angle $\phi_1/f$, this fixes the absolute value of $\phi_1$ via (\ref{eq:dmdensitytd}). We therefore first simulate the ALP dynamics and extract the value of $\mathcal{Z}_0^{\rm osc}$, and only afterwards calculate $\phi_1$ and the corresponding GW signal strength. For the monodromy potential (\ref{eq:monopot}), considered in Sec.~\ref{sec:simple}, we extract this value directly from the lattice simulations, as it was done in~\cite{Berges:2019dgr}. In contrast, in the case of $m_{\rm{final}} \ll m_{a}$, which is investigated in Sec.~\ref{sec:extended}, full numerical simulations become very expensive and, in order to determine $\mathcal{Z}_0^{\rm osc}$, we employ a simplified kinetic description for the dynamics at late times.
\item The energy fraction $\Omega_{\mathrm{GW}, \rm emit}$ is calculated by dividing the energy density in the GWs by $\rho_c = (3/8\pi) M_{\rm Pl}^2 H^2$, where $H$ is the Hubble parameter during the simulation.
\item We use (\ref{nu_today}) and (\ref{om_today}) to transform the simulation variables to today's observables.
\end{itemize}

The simulations were performed on grids with a fixed comoving volume and up to $512^3$ lattice points. We verified that the results are insensitive to variations of the volume and the lattice spacing.

\section{GW production in the simple model}
\label{sec:simple}

We now investigate in more detail the GW production in our lattice simulations for the case of the simple monodromy potential (\ref{eq:monopot}).

The spectrum of ALP fluctuations is conveniently characterized in terms of conformal field and time variables, $d\tau = dt/a$, $\hat \phi_c = \hat \phi a$. The occupation numbers can be then defined as~\cite{Berges:2008wm, Tranberg:2008tg}
\beq
\label{occupnum}
n(t(\tau),\mathbf{p})+1/2 = \sqrt{F_{\mathrm{c}}(\tau,\tau',\mathbf{p}) \partial_{\tau} \partial_{\tau'} F_{\mathrm{c}}(\tau,\tau',\mathbf{p})} |_{\tau'=\tau},
\eeq
where $$F_c(t,t',\mathbf{p}) = \frac{1}{2V} \langle \{ \hat \phi_{c}(t, \mathbf{p}), \hat \phi^{\dagger}_{c}(t', \mathbf{p})  \} - \phi_c(t) \phi_c(t') V \delta_{\rm p 0},\: \: \: \:\: \: \: \: \: \: \: \: \: \: \phi_c(t) =  \langle \hat \phi_c(t, \mathbf{x}) \rangle,$$ are the connected anti-symmetric two-point function and the one-point functions, respectively. This definition is valid for a homogeneous quantum system~\cite{Berges:2015kfa}, such as the one we consider. The large occupation numbers allow to compute the expectation values as classical-statistical averages and ergodicity enables replacing them by volume averages~\cite{Berges:2013lsa}.

\begin{figure}[!t]
	\centering
	\includegraphics[height=0.29\textheight]{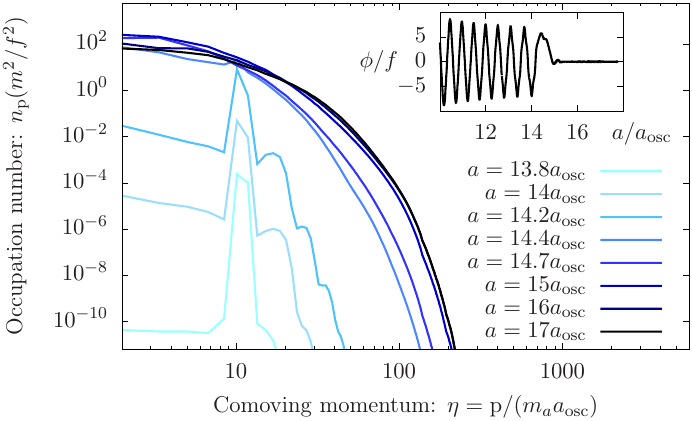}
	\caption{Several snapshots of ALP occupation numbers against their comoving momenta during the fragmentation of the field. Different colors, changing from light to dark blue, correspond to different scale factors in the range $a/a_{\rm osc} = 13.8-17$. The inset shows the evolution of the background field $\phi/f$ at these scale factors. We employ $\phi_1/f=200$, $\kappa=3$ and $H_I/f = 10^{-10}$.}
	\label{spec_alp_att}
\end{figure}

\begin{figure}[!t]
	\centering
	\includegraphics[height=0.289\textheight]{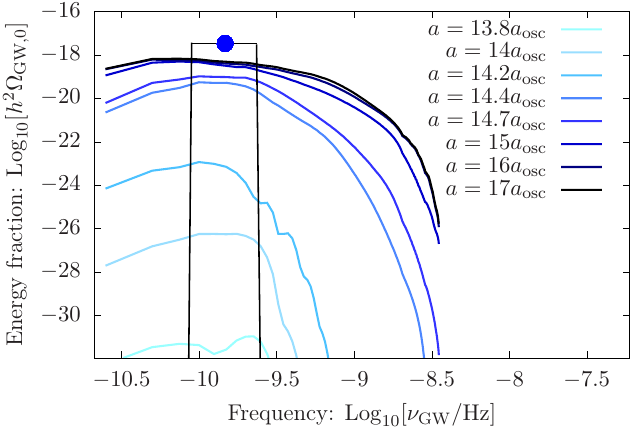}
	\caption{Several snapshots of the fractional GW spectrum, transformed to today's variables, during the fragmentation of the field. Different colors correspond to different scale factors, same as in Fig.~\ref{spec_alp_att}. The blue circle corresponds to the analytical estimate based on \eqref{signal_str_form} and the black rectangle demonstrates the logarithmic width of $\beta=1$, used in that estimate. We employ $\phi_1/f=200$, $\kappa=3$ and $H_I/f = 10^{-10}$, as well as set $m_a = 10^{-16} \rm{eV}$.}
	\label{spec_gw_att}
\end{figure}

Several snapshots of ALP occupation numbers, extracted from numerical simulations, are shown in Fig.~\ref{spec_alp_att} against their comoving momenta, for a particular set of the parameters $\phi_1/f=200$, $\kappa=3$ and $H_I/f = 10^{-10}$. Different colors correspond to different scale factors in the range $a/a_{\rm osc} = 13.8-17$, which is when fragmentation takes place. The inset shows the collapse of the background field $\phi/f$ (the one-point function) at these scale factors. Fig.~\ref{spec_gw_att} shows snapshots of the produced GW spectrum $\mathrm{Log}_{10}[h^2 \Omega_{\mathrm{GW}, 0}]$, as a function of the logarithmic frequency, at the same scale factors as in the previous figure. Here we used $m_a=10^{-16}\rm {eV}$ for the ALP mass. All quantities are transformed to today's values and we used $h=0.68$~\cite{Aghanim:2018eyx}. 

As one can observe in these figures, the two spectra are closely linked to each other. As the GW spectrum is sourced by the fluctuations of the ALP field,
its growth requires large ALP fluctuations.
The resonant growth of ALP fluctuations is accompanied by GW production at momenta of the same order. The broadening of the ALP spectrum due to nonlinear effects is reflected in an analogous broadening of the GW spectrum. At later times the evolution of both spectra slows down, which is due to the interaction rates becoming smaller compared to the expansion rate. We discuss this last aspect in more detail in the next section.

We also compare the GW spectrum at its peak strength, which is acquired during the collapse of the background field, with our analytical estimates from the previous section.  To this end we insert $\nu_{\star} \approx 1.5 \times 10^{-10}\rm Hz$, which corresponds to $\eta_{\star} \approx 10$, into  (\ref{signal_str_form}). We use the numerically extracted value $Z_{0}^{\rm emit} \approx 0.6$ (for this particular set of parameters), where we identify $a_{\rm emit}\approx 15 a_{\rm osc}$. We also set $\alpha = \beta = 1$. The obtained estimate is indicated by the blue circle in Fig.~\ref{spec_gw_att}. As can be seen it matches reasonably well with the numerical calculation, although it is somewhat higher compared to the numerical result. The latter can be understood from the fact that $\beta =1$, which was used in that analytical estimate and would approximately correspond to the black rectangle in Fig.~\ref{spec_gw_att}, is smaller than the actual logarithmic width of the spectrum, as can be seen in that figure. 

\begin{figure}[!t]
 	\centering
 	\includegraphics[height=0.29\textheight]{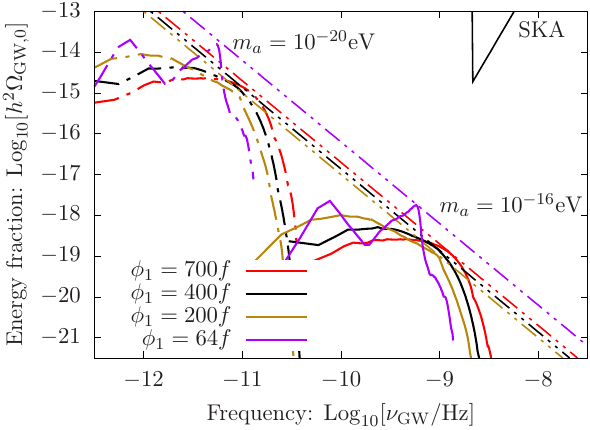}
 	\caption{The parameter space for the numerically extracted GW spectra from fragmentation for the potential of (\ref{eq:monopot}) for different misalignment angles $\phi_1/f$. The dashed lines indicate the numerically obtained envelopes of the signals scaling as in~\eqref{signal_str_form}.  We employ $\kappa=3$ and $H_I/f=10^{-10}$. For comparison we show the SKA sensitivity~\cite{Smits:2008cf,Moore:2014lga}.}
 	\label{gw1}
\end{figure}

Having extracted the late-time (frozen) GW spectra, in Fig.~\ref{gw1} we plot them for different ALP masses and misalignment angles, thereby obtaining the parameter space spanned by the GW signal, which we indicate by the dashed slopes. As can be seen, larger misalignment angles lead to a gravitational production at higher frequencies, however the overall parameter space spanned by the signal is not affected by this dramatically. We have also performed simulations with a larger value of $\kappa=6$, which led to a strength of the gravitational signal very similar to the one from $\kappa=3$.
 
\section{Extended relativistic phase and a stronger GW signal}
\label{sec:extended}

In this section we investigate the fragmentation of ALP DM and the emission of GWs in a more general setting. This provides scenarios in which the gravitational signal is enhanced and can even reach into the range detectable by future experiments.

As it was explained in Sec.~\ref{ssec:summary}, a stronger stochastic GW signal can be produced if $\mathcal{Z}_0^{\rm emit} \ll 1$ (see Eq.~(\ref{signal_str_form})). Such efficient dilution of the ALP energy density after fragmentation can be induced if ALPs exhibit an extended intermediate phase of relativistic dynamics. During this phase the energy density decreases as $a^{-4}$ i.e.~faster than in a situation where the ALPs are nonrelativistic. Such an intermediate phase is present also in the standard scenario, discussed in the previous section, however it is typically short and ends when the characteristic momenta, due to the red-shift, drop below the mass $m_a$. 

\bigskip

The ultrarelativistic phase is extended in the case when the final mass near the bottom of the potential is much smaller than $m_a$. This can happen if the sign in front of the periodic term is negative. Up to an irrelevant constant in the potential this simply corresponds to having a phase shift of $\delta=\pi$ in the periodic part of the potential in Eq.~\eqref{eq:monopot}, 
\beq
U(\phi) = \frac{1}{2}m^2 \phi^2  - \Lambda^4 \Bigl(1-\cos\frac{\phi}{f} \Bigr).
\eeq
Near its minimum the potential has the form
\beq
\label{inverse_cosine_pot2}
U(\phi) = \frac{1}{2} (1-\kappa^2) m^2 \phi^2  + \frac{\lambda}{4!}\phi^4  - \frac{\lambda f^{-2}}{6!}\phi^6 + ...
\eeq
and is convex if $\kappa^2<1$. Here $\lambda = \Lambda^4/f^4$ denotes the quartic coupling. By taking $\kappa \approx 1$, the mass near the bottom of the potential can be made very small. 

\begin{figure}[!t]
	\centering
	\includegraphics[width=0.8\textwidth]{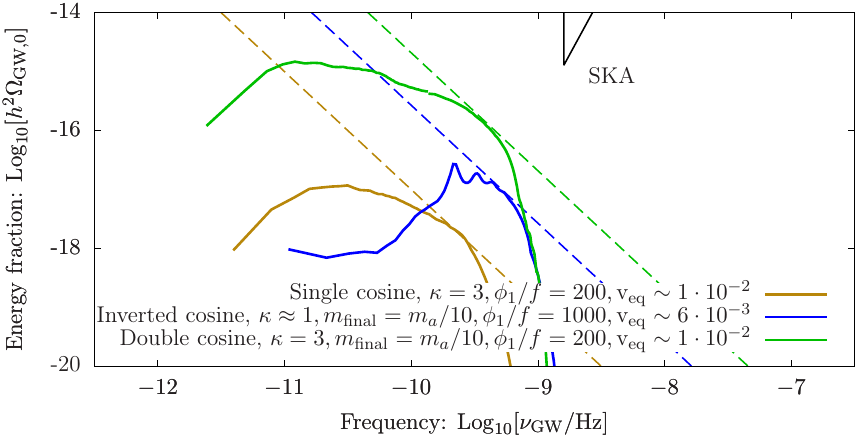}
	\caption{The parameter space for the numerically extracted GW signal from fragmentation for the potential (\ref{eq:monopot}) of the simple model (golden), for the ``inverted single cosine'' potential (\ref{inverse_cosine_pot2}) with $m_{\rm  final} = m_a/10$ (blue), and for the ``double cosine'' potential (\ref{double_cosine_pot1}) with $m_{\rm  final} = m_a/10$ (green). The dashed slopes indicate the bounds for the signal, as in Fig.~\ref{gw1}. We employ $m_a = 10^{-17}\rm eV$, $\kappa=3$ and $H_I/f=10^{-10}$.
	We also show the sensitivity of the SKA pulsar timing array~\cite{Smits:2008cf,Moore:2014lga}.}
	\label{gwk1}
\end{figure}

We have performed numerical simulations for this potential. Unfortunately, in the ``inverted cosine'' potential with a small mass at the bottom the restriction to values $\kappa^2\approx 1$ leads to a rather weak resonance. As a result, strong fragmentation requires relatively large misalignment angles, $\phi_1/f\gtrsim 500$ (see also Fig.~8 in~\cite{Berges:2019dgr}), and occurs at later times, leading to a broader and weaker spectrum. This is illustrated in Fig.~\ref{gwk1}, where the final GW spectra for the simple (gold), ``inverted cosine'' (blue) and ``double cosine'' (green, see the following for details) models are shown for sets of parameters that leads to a strong signal in that model. While the simple model produces the weakest signal, the spectrum from the ``inverted cosine'' is weaker compared to the one from the ``double cosine'' for the same ratio of the final mass to $m_a = m\sqrt{1+\kappa^2}$ and similar velocities at matter-radiation equality.

To remedy this we can consider a ``double cosine'' modulation,
\beq
\label{double_cosine_pot1}
U(\phi) = \frac{1}{2}m^2 \phi^2  + \Lambda^4 \Bigl(1-\cos\frac{\phi}{f} \Bigr)  + r \Lambda^4 \Bigl(1-\cos \frac{2\phi}{f} \Bigr).
\eeq
Near the minimum the potential has the form
\beq
\label{double_cosine_pot2}
U(\phi) = \frac{1}{2} (1+\kappa^2 + 4 r \kappa^2) m^2 \phi^2  - \frac{\lambda(16r+1)}{4!}\phi^4 + \frac{\lambda f^{-2}}{6!} (1+64r)\phi^6 -...
\eeq
and, if the value of $r$ very slightly exceeds $-(1+\kappa^2)/4\kappa^2$ (for large values of $\kappa$ this corresponds to $r\approx-1/4$), the mass near the bottom is again very small. 

In the case of the ``double cosine'' potential, the parameter $\kappa$, which determines the size of the wiggles and, therefore, the strength of the resonance, is not restricted to $\kappa\approx1$ and can be large. This is in contrast to the ``inverted single cosine'' modulation, where $\kappa\approx1$ is necessary to have a small mass at the bottom of the potential. Because of this higher generality we focus on this second possibility. 

\subsection{The role of the small mass and repulsive self-interactions}
\label{sec:att_rep}
The tuning of the bare mass near the bottom of the potential does not qualitatively change the early dynamics, including the amplification of fluctuations and the decrease of the background field amplitude\footnote{In particular, since the second cosine has a smaller amplitude then the first one.}. The modification however plays an important role after the collapse of the background oscillations i.e.~when the field settles near $\phi=0$. A smaller mass means that it takes a longer time until the self-interaction becomes unimportant due to the dilution. In particular, the spectrum freezes much later in this case. This makes the analysis based on lattice simulations harder due to the larger range of scales involved and longer required simulation times.

Let us now consider the qualitative behavior of the evolution at relatively late times and small fields, and see which terms are relevant. The sextic and higher-order self-interactions in (\ref{inverse_cosine_pot2}) or (\ref{double_cosine_pot2}) are sub-dominant and can be neglected, already when $\langle \phi^2\rangle/f^2 \lesssim 1$. Importantly, the small value of the bare mass near the bottom however becomes relevant only when the typical field values become sufficiently small. Indeed, writing the effective mass as 
\beq
M^2 = m_{\mathrm{final}}^2 + \frac{\lambda}{2} \langle \phi^2 \rangle,
\label{eq:eff_mass}
\eeq
it follows that the transition to the ``mass-dominated'' regime occurs when $ \lambda \langle \phi^2 \rangle \sim m_{\rm final}^2 $ or, equivalently, 
\beq
\Bigl( \frac{\kappa^2+1}{\kappa^2} \Bigr) \frac{m_{\rm final}^2}{m_a^2} \sim \frac{\langle \phi^2 \rangle}{f^2}
\label{transition_nonrel}
\eeq
is satisfied\footnote{For $\kappa \gtrsim 1$ the term in the brackets (\ref{transition_nonrel}) is of order one.}. In the case of interest to us, we have $m_{\rm final}^2/m_a^2 \ll 1$. Therefore, for $m_{\rm final}^2/m_a^2\lesssim \langle \phi^2\rangle/f^2 \lesssim 1$ we have an extended intermediate stage of the dynamics, during which the potential can be well approximated as a massless (conformal) quartic potential.

In parallel to the already discussed enhanced GW production due to the longer relativistic phase, there is another interesting qualitative difference compared to the original ``single cosine'' potential (\ref{eq:monopot}) with $\delta=0$. In the situation described above the leading quartic self-interactions near the bottom of the potential are now repulsive\footnote{In the second case $r$ is at most approximately $-1/4$ in \eqref{double_cosine_pot2}.}, in contrast to the standard case, in which they are attractive. Repulsive interactions lead to the build-up of a homogeneous and long-lived condensate after the fragmentation via an inverse particle cascade in momentum space~\cite{Orioli:2015dxa, Berges:2017ldx}.

\begin{figure}[!t]
	\centering
	\includegraphics[width=0.63\textwidth]{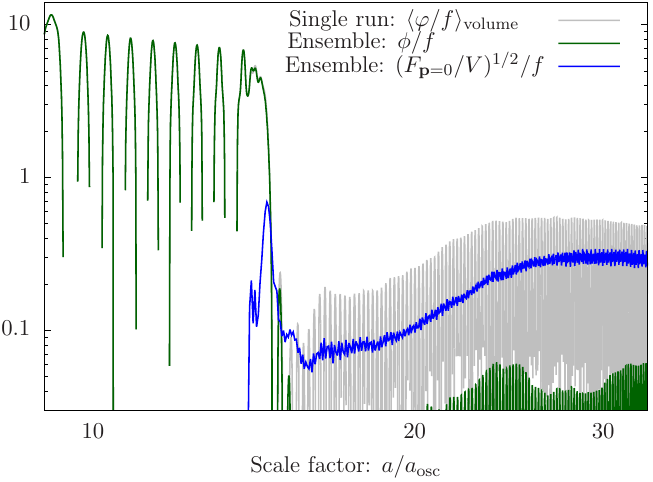}
	\caption{The collapse of the ALP field and condensation in the presence of repulsive self-interactions. The green line corresponds to the one-point function, the blue line to the zero-momentum mode of the two-point function and the gray line represents the volume average of the field within one classical simulation. We employ $\phi_1/f=200$, $\kappa=3, H_I/f = 10^{-10}$ and $m_{\rm final} = 10^{-3}m_a$.}
	\label{cond_build}
\end{figure}

While the initial homogeneous field and the later condensate are very similar, it is worth emphasizing their difference from the point of view of classical-statistical field theory. The first one corresponds to a nonvanishing field one-point function, $\phi(t)$. After the collapse however this expectation value vanishes and the build-up of the condensate instead corresponds to the growth of the zero-momentum mode of the two-point function $F(t, t, \mathrm{p}) |_{\mathbf{p}=0}$. While in each simulation of the classical-statistical ensemble a homogeneous field component emerges as a result of condensation, leading to a nonvanishing volume average of the field, their phases become uncorrelated after the collapse, resulting in a vanishing one-point function. This loss of ergodicity is due to the emergence of a long-range order in the system. Both $\phi(t)$ and $F(t, t, \mathrm{p}) |_{\mathbf{p}=0}$, extracted from classical-statistical simulations, as well as the volume averaged field from a single classical simulation are shown in Fig.~\ref{cond_build}, as functions of the scale factor, for $m_{\rm final} = 10^{-3}m_a$.

In principle one could worry that the existence of a homogeneous condensate is detrimental to the production of GWs. However, as can be inferred from Fig.~\ref{cond_build}, the condensate usually only comprises a quite small fraction of the energy density right after fragmentation and the production of GWs is still significantly enhanced compared to the ``single cosine'' case.

\subsection{GW production}
\label{sseq:GWprod_rep}

\begin{figure}[!t]
	\centering
	\includegraphics[height=0.3\textheight]{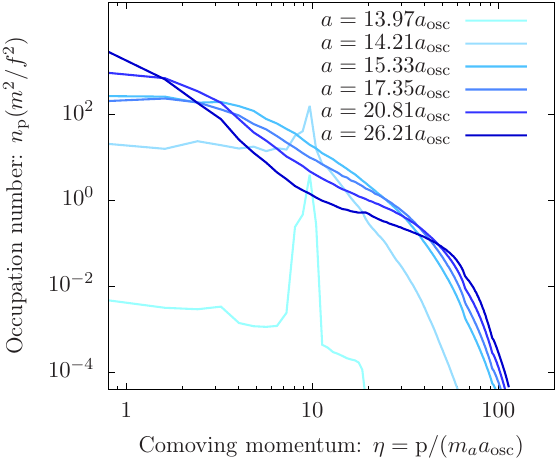}
	\hspace*{0.6cm}
	\includegraphics[height=0.3\textheight]{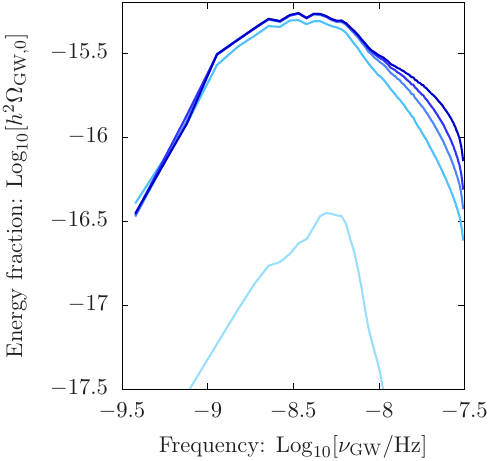}
	\caption{Snapshots of ALP occupation numbers (left) and fractional GW spectra (right) in the presence of repulsive self-interactions. Colors from light to dark blue correspond to scale factors in the range $a/a_{\rm osc}=14-26 $. We employ $\phi_1/f=200$, $\kappa=3, H_I/f = 10^{-10}$ and $m_{\rm final} = 10^{-3}m_a$.}
	\label{spec_rep}
\end{figure}

We now demonstrate the production of GWs during and after fragmentation, similar to how it was done in Sec.~\ref{sec:simple}. In the left panel of Fig.~\ref{spec_rep} we plot several snapshots of ALP occupation numbers at different scale factors, as functions of the comoving momentum. The snapshots of the fractional GW spectrum at the same scale factors are shown in the right panel of Fig.~\ref{spec_rep}. Here the strength of the signal is obtained by estimating the value of $\mathcal{Z}_0^{\rm osc}$, which is explained in the next subsection. One observes two main differences compared to Figs.~\ref{spec_alp_att} and~\ref{spec_gw_att}:
\begin{itemize}
	\item{There is a more regular direct cascade, compared to the attractive case, which is accompanied by GW radiation at corresponding momenta. The cascade transports energy to higher momenta and, after a some transient time, is governed by a self-similar time evolution~\cite{Micha:2004bv, Berges:2017ldx}, i.e.
	\beq
	n(\tau,\mathrm{p})= \Bigl( \frac{\tau}{\tau_S} \Bigr)^{\alpha}n_S\Bigl(  \Bigl( \frac{\tau}{\tau_S} \Bigr)^\beta \mathrm{p} \Bigr),
	\eeq
	where $\tau$ is the conformal time, $n_S$ denotes the spectrum at some reference time and the scaling exponents are found to be $\beta \approx -1/5$ and $\alpha \approx 4\beta$~\cite{Micha:2004bv}. Due to the energy cascade the characteristic comoving momentum, which we denote by $\bar{\mathrm{p}}$ grows as $\bar{\mathrm{p}}(\tau) \propto \tau^{-\beta}$.}
	\item{In addition to the direct cascade, an inverse cascade emerges transporting particle number to low momenta. The evolution of the inverse cascade also becomes self-similar, with exponents $\beta\approx 1/2$ and $\alpha \approx 3\beta$~\cite{Berges:2014xea, Berges:2017ldx}. As can be seen, at least in the model at hand the inverse cascade does not produce significant amounts of GWs.}
\end{itemize}

Remarkably, the fragmentation of the ALP field is accompanied by a gravitational signal of a similar strength as in the attractive case. In other words, even though a quasi-homogeneous field is re-created after fragmentation, the intermediate phase and the remaining fragmented part are sufficient to produce a strong signal.
Therefore, we can fully benefit from the longer relativistic evolution to achieve a stronger GW signal today.

\subsection{Extrapolation of the dynamics to late times}
\label{extrapol}

A smaller mass near the bottom of the potential requires a longer simulation of the dynamics, which becomes increasingly difficult to perform on a lattice. A rough way to calculate the prefactor $Z^{\rm osc}_0$ would be to use the latest available ALP spectrum from the simulation for later times of the dynamics. This however neglects the direct cascade of the distribution. In this work we take into account the self-similar evolution of the direct cascade and extrapolate the evolution of the spectrum by solving the kinetic equation for the characteristic momentum scale $\bar{\mathrm{p}(a)}$. We also keep track of the occupation number at the characteristic momentum  $\bar{n}$ and the condensate amplitude $\langle \phi^2 \rangle$ as functions of the scale factor.
The details of our method for the extrapolation are described in the appendix. Below we directly present the results.

\begin{figure}[!t]
	\centering
	\includegraphics[width=0.47\textwidth]{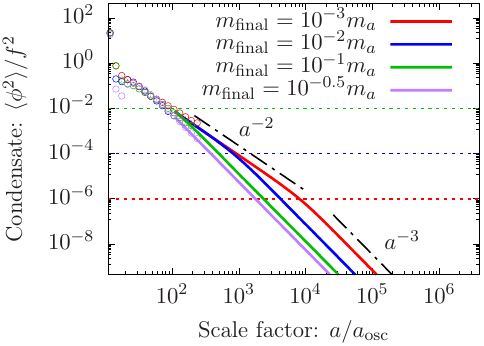}
	\hspace{0.7cm}
	\includegraphics[width=0.47\textwidth]{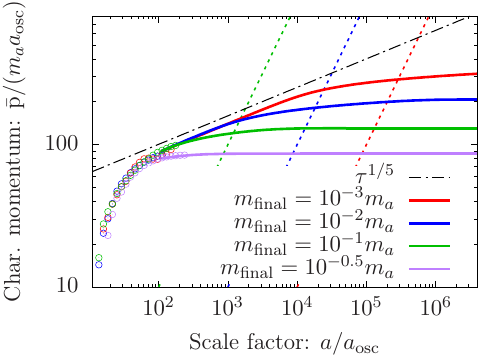}
	\caption{The amplitude of the condensate $\langle \phi^2 \rangle/f^2$ (left) and the characteristic comoving momentum  $\bar{\mathrm{p}}/(m_a a_{\rm osc})$ (right) as functions of the scale factor for different mass ratios $m_{\rm final}/m_a$, indicated by different colors. The points are from lattice simulations and the solid lines correspond to the extrapolation. In the left panel the dotted lines show the corresponding ratios $m_{\rm final}^2/m_a^2$ for each color. They approximately separate the relativistic ($\propto a^{-2}$) and the nonrelativistic ($\propto a^{-3}$) regimes of the condensate dynamics, as explained in the main text. In the right panel the dotted lines of different colors denote $(m_{\rm final}/m_a) (a/a_{\rm osc})$, such that their intersection with the corresponding solid line is at $\bar{\rm p}_{\rm phys} \approx m_{\rm final}$. The dashed black line indicates the $\tau^{1/5}$ growth of $\bar{\rm p}$, observed at early times in that plot. We employ $\phi_1/f=200$, $\kappa=3$ and $H_I/f = 10^{-10}$. }
	\label{extrapol_1}
\end{figure}

In the left panel of Fig.~\ref{extrapol_1} the extrapolated condensate amplitude $\langle \phi^2 \rangle/f^2$ is shown with solid lines as a function of the scale factor. The points at early times are from the lattice simulations. Different colors correspond to different values of $m_{\rm final}^2/m_a^2$. These values  are also shown in the plot with the dotted lines in corresponding colors. One observes the transition from the quartic-dominated relativistic regime, characterized by an approximately constant $\langle \phi_c^2 \rangle$, to the mass-dominated nonrelativistic regime during which $\langle \phi^2 \rangle \propto a^{-3}$. As it was mentioned in Sec.~\ref{sec:att_rep}, the transition between these regimes happens approximately when \eqref{transition_nonrel} is satisfied i.e.~when the solid line intersects the dotted line. 

In the right panel of Fig.~\ref{extrapol_1} the extrapolated evolution of the characteristic comoving momentum $\bar{\mathrm{p}}/(m_a a_{\rm osc})$ is shown with solid lines for the same values of the mass ratios as in the left panel. The points of corresponding color at early times are from the lattice simulation. One indeed observes the $\propto \tau^{1/5}$ growth at early times. The growth slows down once the bottom mass becomes important i.e.~when the transition to the nonrelativistic regime in the left panel occurs. The characteristic momenta at this point are still relativistic. This can be seen by comparing the solid lines to the dotted lines of corresponding colors, which denote $(m_{\rm final}/m_a) (a/a_{\rm osc})$. The intersection of these two lines corresponds to $$\bar{\rm p}/a \approx m_{\rm final},$$ i.e.~the characteristic high momentum modes becoming nonrelativistic.

Our extrapolation allows to keep track of the equation of state of the system. We do this by re-scaling the momentum occupation numbers according to $\bar{\rm p}(a)$ and $\bar{n}(a)$, and calculating the energy and the pressure of both components of the system. The equation of state can be written as
\beq
w = \frac{p_c}{\rho_c} = \frac{p_{\phi,c} +  V^{-1}\int_{\mathbf{p}} n_{\rm p} ({\rm p}^2/3\omega_{\mathrm{p}} )  }{\rho_{\phi,c} +V^{-1} \int_{\mathbf{p}} n_{\rm p} \omega_{\mathrm{p}} },
\eeq
where for simplicity we work in conformal variables. 

\begin{figure}[!t]
	\centering
	\includegraphics[width=0.47\textwidth]{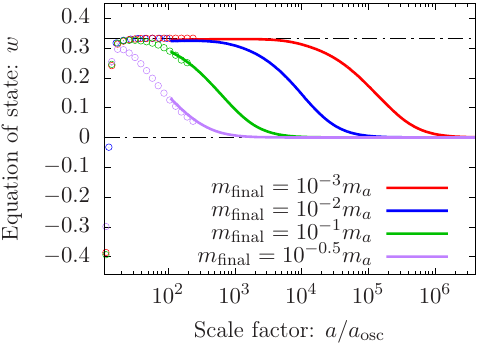}
	\caption{The equation of state parameter $w = p/\rho$ as a function of the scale factor for different mass ratios $m_{\rm final}/m_a$, indicated by different colors. The points are from lattice simulations and the solid lines correspond to the extrapolation. We employ the same parameters as in Fig.~\ref{extrapol_1}.}
	\label{extrapol_2}
\end{figure}

Calculated in this way $w(a)$ is shown in Fig.~\ref{extrapol_2} for the same values of the mass ratios as in Fig.~\ref{extrapol_1}. The points of corresponding color at early times are from the lattice simulation. We average $w$ over an oscillation period to get rid of its oscillations. The plot demonstrates the expectation that the ALPs exhibit an intermediate phase with $w  = 1/3$. The transition to $w=0$ completes when the high momentum modes become nonrelativistic. Integrating $w(a)$ gives the values of $\mathcal{Z}^{\rm osc}_0$, used in calculating the GW spectra.

We note that, in particular for the green and purple lines in Fig.~\ref{extrapol_2}, corresponding to $m_{\rm{final}}=10^{-1}m_{a}$ and $m_{\rm{final}}=10^{-0.5}m_{a}$ respectively, we have a sizeable overlap between the numerical data and the analytic extrapolation where we see good agreement. While this is in part due to the involved fitting of parameters between the simulation and the analytical approximation we nevertheless take this as an indication of the validity of our approximation.

\subsection{Constraints from structure formation}

After the collapse the ALP field splits into a quasi-homogeneous condensate and high-momentum particles. Initially the latter take most of the energy of the system, more than $90\%$. At the end of nonlinear dynamics this fraction however decreases significantly, and the two contributions are of roughly the same order of magnitude. The reason is that in the late phase of the dynamics the condensate dilutes as $\propto a^{-3}$, whereas the relativistic high-momentum particles feature a $\propto a^{-4}$ behavior. 
Nevertheless, the fraction of relatively high momentum particles is sizable.
In this section we therefore discuss the constraints for successful structure formation for both types of field configurations.

\paragraph{Constraints on homogeneous oscillations:} The authors of~\cite{Cembranos:2018ulm} derived constraints from linear structure formation (at scales $\mathrm{k} \lesssim 0.2 h/{\rm Mpc}$) for DM in the form of a homogeneously oscillating scalar field with a potential $U(\phi) = m^2 \phi^2 /2 + \lambda \phi^4/4$, where $\lambda>0$ (which is exactly the case of main interest to us). They obtained the following bound on the mass and the coupling of the field,
\beq
\log_{10}( \lambda ) < -91.86 + 4 \log_{10}\Bigl( \frac{m}{10^{-22}\rm eV} \Bigr).
\eeq

For sufficiently small masses at the bottom, the potential (\ref{double_cosine_pot1}) is very close to a quartic one at late times. We can thus directly apply the above bound to the considered set-up. Substituting $$m \rightarrow m_{\rm final}, \: \: \: \: \: \: \: \: \: \: \: \: \: \: \: \: \: \: \: \: \: \: \: \: \: \: \: \lambda \rightarrow -\frac{\lambda (16r + 1)}{3!} \approx \frac{\lambda}{2} \approx \frac{m_a^2}{2 f^2},$$ one arrives at
\beq
\label{contr_field}
2 \log_{10}\Bigl( \frac{m_a}{m_{\rm final}} \Bigr) + 1.56 < 2 \log_{10}\Bigl( \frac{m_{\rm final}}{10^{-22}\rm eV} \Bigr) + 2 \log_{10}\Bigl( \frac{f}{10^{14}\rm GeV} \Bigr).
\eeq
For given values of the masses this represents a lower bound on the decay constant (and for fixed $\phi_1/f$ also on the misalignment field value $\phi_1$). For the situations indicated by solid lines in our summary Fig.~\ref{gw_main} this bound implies $f > 6\times10^{11} \rm GeV$, while for the ones indicated by dashed lines $f>6\times10^{10} \rm GeV$. The actual values of the decay constant are about $(2-3)\times10^{14} \rm GeV$ for the solid lines and about $(1-2)\times10^{14} \rm GeV$ for the dashed lines. The constraints are thus satisfied.

We are not aware of similar constraints from nonlinear structure formation, which are, however, expected to be stronger.

\paragraph{Constraints on typical velocities:} The typical velocities of ALPs at matter-radiation equality are related to their characteristic comoving momentum $\bar{\rm p}$ via (\ref{exp_for_v}). A first simple constraint on the velocities can be obtained by demanding that the ALPs are nonrelativistic ($\bar{\mathrm{v}} \lesssim 1$) at the time when the structures on linear scales enter the horizon, $\mathrm{k}/a \sim H$ with $\mathrm{k} \lesssim 0.2 h/{\rm Mpc}$~\cite{Colombi:1995ze, Bode:2000gq, 10.1111/j.1365-2966.2004.07358.x}, which leads to $\mathrm{v_{eq}} \lesssim 10^{-1}$. (The turnover of the power spectrum at the scale $\mathrm{k} \sim 0.02 h/{\rm Mpc}$ corresponds to the horizon entry at matter-radiation equality). Stronger constraints on the velocities arise from the Lyman-$\alpha$ forest data on nonlinear structure formation. For the smallest observed structures on scales $ \mathrm{k} \sim 5 h/{\rm Mpc}$ the bound becomes $\mathrm{v_{eq}} \lesssim 10^{-3}$ (see, e.g.,~\cite{Bode:2000gq, Kunz:2016yqy}). Similar values can be obtained from translating (cf., e.g.~\cite{Berges:2019dgr}) limits on the mass of thermally produced warm dark matter candidates investigated, e.g. in~\cite{Viel:2005qj, Viel:2007mv, Baur:2015jsy}. More detailed numerical simulations would be required to obtain more precise constraints that take into account the precise, non-thermal, velocity distribution of ALPs.

We estimated numerically the typical velocities at matter-radiation equality from the late-time asymptotic values of the extrapolated dynamics of $\bar{\rm p}$, shown in the right panel of Fig.~\ref{extrapol_1}, using (\ref{exp_for_v}). For the solid curves ($h^2 \Omega_{\mathrm{GW}} \sim 10^{-15} - 10^{-14}$) in Fig.~\ref{gw_main} these typical velocities are $\sim 10^{-2}$, whereas for the dotted lines $\mathrm{v}_{\rm eq} \sim 10^{-3}-10^{-2}$. The constraints from linear structure formation are clearly satisfied in this case. The situation for nonlinear structure formation is less clear and the particles might turn out to be somewhat warmer. On the other hand it is important to realize that a significant fraction of the energy density of ALPs is in the form of a homogeneous field, which is expected to decrease the ALP pressure and soften the constraints. Answering this question requires further investigation.

Fig.~\ref{gw_main} also illustrates that an appropriate (tuned) choice of the masses allows to shift the signal towards larger frequencies while keeping both the typical velocities at matter-radiation equality and the signal strength approximately unchanged.

\subsection{Resulting gravitational wave spectra}
\label{ssec:resulting_spectra}

\begin{figure}[!t]
	\centering
	\includegraphics[width=0.98\textwidth]{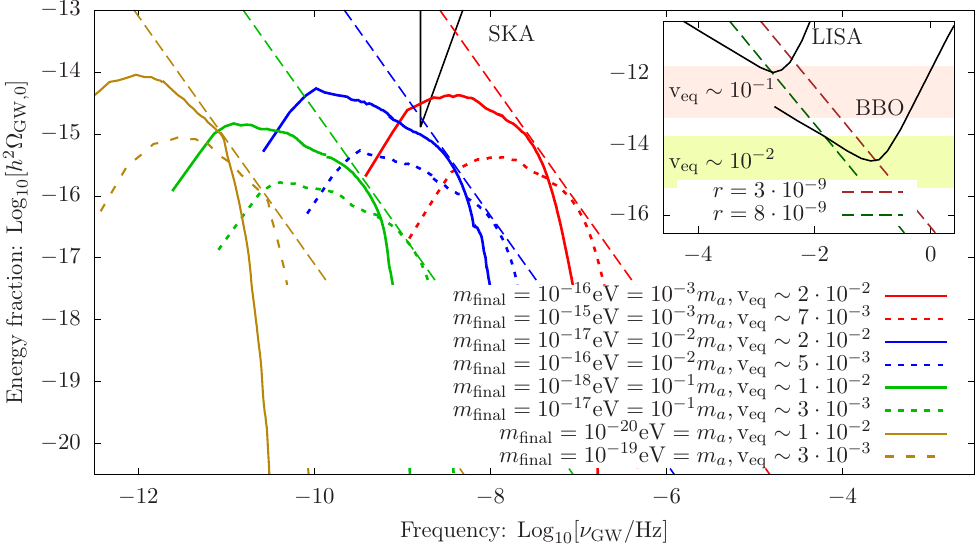}
	\caption{Gravitational wave spectra in models with an extended relativistic phase. The spectra are shown for several values of the mass with different colors corresponding to different mass ratios $r = m_{\rm final}/m_a$. The golden color corresponds to the nonmodified potential (\ref{eq:monopot}) with $\delta=0$ and attractive self-interactions, also shown in Fig.~\ref{gw1}. The dashed slopes indicate the envelopes of the signal strength  as in Fig.~\ref{gw1}. The inset shows analytical estimates for the signal envelopes at smaller values of $r$, as explained in section~\ref{ssec:resulting_spectra}. The sensitivities of relevant GW experiments are indicated with solid black lines according to~\cite{Moore:2014lga} and the estimated typical velocities at matter-radiation equality are stated. We employ $\phi_1/f=200, \kappa=3$ and $H_I/f=10^{-10}$.}
	\label{gw_main}
\end{figure}

Here we present the main results obtained from our numerical setup. These are summarized in Fig.~\ref{gw_main}, where the gravitational spectra, extracted from the latest times of the numerical simulations are shown. At these times the ALP field has already fragmented and most of the GWs have been produced much earlier. The chosen parameters are $\kappa=3$, $H_I/f = 10^{-10}$ and $\phi_1/f=200$ as benchmark parameters for the figure.

The gold curves correspond to the simple potential (\ref{eq:monopot}), such that $m_{\rm final}=m_a$, and different types of lines correspond to different values of the mass. The corresponding parameter space for the gravitational signal is indicated by the envelope line of the same color. The remaining curves in Fig.~\ref{gw_main} correspond to the late-time spectra of GWs in the case when $m_{\rm final} \ll m_a$. Each color corresponds to a particular ratio $m_{\rm final}/m_a$. One indeed observes the increasing range of the signal in the parameter space as this ratio is being decreased, consistent with the expectation from (\ref{signal_str_form}). Remarkably, already for $m_{\rm final} = 10^{-3} m_a$ the range of the signal overlaps with the planned sensitivity of pulsar timing arrays, i.e. the Square Kilometre Array (SKA)~\cite{Smits:2008cf,Moore:2014lga}. In particular, the solid red curve corresponds to $$ m_{\rm final}  = 10^{-16} \mathrm{eV}, \: \: \: \: \: r = 10^{-3}\: \: \: \: \:  \: \: \: \: \: \: \: \: \: \:(\text{SKA}).$$

It is possible to extend the signal range even further, up to the future sensitivities of space-based interferometers, although this requires an even stronger tuning of the mass near the bottom. In particular, the Big Bang Observer (BBO)~\cite{Crowder:2005nr} and the Laser Interferometer Space Antenna (LISA)~\cite{Audley:2017drz}) are most sensitive to signals peaked at around $\nu \sim 10^{-1}\rm Hz$ and $\nu \sim 10^{-3}\rm Hz$, respectively. From the analytical estimate (\ref{nu_today}), it follows (using $\eta_{\star}\sim 10$ as before) that the ALP masses $m_a$ most relevant for these detector are 
$$m_a \sim 40 \, \mathrm{eV}  \text{ for BBO, } \: \: \: \: \: \: \: \: \:  m_a \sim  10^{-2}\,  \rm eV  \text{ for LISA}.$$
From the planned sensitivity of BBO, $\Omega_{\star} \sim 10^{-15}$, and that of LISA, $\Omega_{\star} \sim 10^{-12}$, the corresponding enhancement factor $\mathcal{Z}^{\rm emit}_{0}$ can be estimated using (\ref{signal_str_form}) to be $\mathcal{Z}^{\rm emit}_{0} \sim 2\times 10^{-11}$ and $\mathcal{Z}^{\rm emit}_{0} \sim 6\times10^{-11}$, respectively. The final step is to determine the mass ratios $r = m_{\rm final}/m_a$ which lead to such values of the enhancement factor. Here we estimate the dependence $\mathcal{Z}^{\rm emit}_0 (r)$ analytically. To this end we note that, as it follows from the definition~(\ref{eq:Z}), $\mathcal{Z}^{\rm emit}_{0}(r) \propto a_{\mathrm{nr}}^{-1}(r)$, where $a_{\mathrm{nr}}(r)$ is the scale factor at which the transition to the nonrelativistic regime takes place for given $r$. The latter can be estimated from the condition
 $$ m_{\rm final}(r)   = m_a r  \sim \frac{\bar{\mathrm{p}}(a_{\mathrm{nr}}(r))}{a_{\mathrm{nr}}(r)}  \propto a_{\mathrm{nr}}^{-4/5}(r), $$ 
 where we used the form of the self-similar evolution $\bar{\mathrm{p}}(a) \propto \tau^{1/5}(a) \propto a^{1/5}$ from Sec.~\ref{sseq:GWprod_rep} (see also the appendix). This implies a simple approximate dependence $\mathcal{Z}^{\rm emit}_0(r) \propto r^{5/4}$, and, using the already calculated values of $\mathcal{Z}^{\rm emit}_0$ for $r\geq 10^{-3}$ we arrive at the following estimates for the relevant masses that can lead to a detectable signal, 
 $$ m_{\rm final}  \sim 10^{-8} \mathrm{eV}, \: \: r \sim 3\times 10^{-9}\: (\text{BBO}),\: \: \: \:  \: \: \: \: \: \: \: \: \: \:  m_{\rm final}  \sim 10^{-11} \mathrm{eV},  \: \: r \sim 6\times 10^{-9}\:(\text{LISA}).$$
The signal boundaries based on these two values of $r$ are shown in the inset of Fig.~\ref{gw_main}, where also the sensitivity curves of the corresponding detectors are shown with black continuous lines~\cite{Moore:2014lga}. The anticipated bands for the values of typical velocities at matter-radiation equality $\mathrm{v}_{\rm eq}\sim 10^{-1}$ and $\mathrm{v}_{\rm eq}\sim 10^{-2}$ are indicated by the pink and yellow colors. Note that although less fine-tuning is required for the LISA case, the typical velocities near matter-radiation equality are quite large and, therefore, are likely to cause trouble with nonlinear structure formation.

\section{Conclusions}
\label{sec:conc}

The development of GW detection experiments opens new possibilities for exploring the physics of the early universe and, in particular, getting insights into the nature of dark matter. In this work we described how such experiments can detect signatures of the nonperturbative dynamics of ALP DM shortly after its production via the vacuum misalignment mechanism. Such dynamics is common for ALPs with a broken discrete shift symmetry, e.g., in the presence of a monodromy. 

We have estimated the stochastic GW background both analytically as well as numerically. Despite the strong constraints on the signal, arising from the requirement of not over-producing the dark matter, the signal from fragmentation can be sufficiently strong to be observed by GW experiments. This can happen if ALPs exhibit an extended relativistic phase of dynamics after fragmentation, a scenario which we studied in detail in this work, by assuming a small mass near the bottom of the potential. 

Our main findings are summarized in Fig.~\ref{gw_main}, which contains the relevant GW spectra. Potentially a signal can be detectable by pulsar timing arrays, such as the one based on SKA for ALPs with $m_{\rm final} \sim 10^{-16} \rm{eV}$. Here a fine-tuning of the order $10^{-3}$ of the mass at the bottom of the potential is required. An even more severe fine tuning of the order of $10^{-10}$ would bring the signal into the sensitivity range of space-based interferometers such as BBO ($m_{\rm final} = 10^{-8} \rm{eV}$) or LISA ($m_{\rm final} = 10^{-11} \rm{eV}$). We have also confirmed that the constraints from linear structure formation are satisfied in these cases. The situation with  nonlinear structure formation is less clear and requires further investigation.  In particular, it would be interesting to understand the role of the quasi-homogeneous condensate, which emerges in the presence of repulsive self-interactions, on structure formation, as well as find possible observational signatures.

\section{Acknowledgements}
We would like to thank J\"urgen Berges for many insights as well as collaboration on related work. We are also grateful to Gonzalo Alonso-\'Alvarez, Ippei Obata,  Wolfram Ratzinger, Pedro Schwaller and Pablo Soler for helpful discussions. Some of the numerical calculations were performed on the computational resource bwUniCluster, funded by the Ministry of Science, Research and the Arts Baden-W\"urttemberg and the Universities of the State of Baden-W\"urttemberg, within the framework program Baden-W\"urttemberg high performance computing (bwHPC).

\appendix

\section{Appendix: Simplified kinetic description}
\label{app}

In this appendix we describe a simplified analysis of the late-time dynamics, based on the kinetic description, which is then  used to obtain estimates for the energy loss factor $\mathcal{Z}_0^{\rm osc}$ and for the typical velocities at matter-radiation equality. We consider the scenario of a small mass near the bottom of the potential, which leads to an extended phase of nonlinear dynamics and strong GW production.

After the fragmentation of the ALP field, the momentum distribution function, in the presence of repulsive self-interactions, splits into two cascades, separated approximately at the momentum equal to the the mass (see Fig.~\ref{spec_rep}). This allows us to characterize the field dynamics in terms of a few quantities describing both momentum regions, as we describe below.

At low momenta, the final result of the inverse cascade is the emergence of a homogeneous condensate, which carries most of the particle number. In the lattice simulations this happens after a finite time that increases with the volume. In other words, the condensate does not become homogeneous across the whole universe within a finite amount of time. Nevertheless, the typical scale of inhomogeneities in the infrared region is from the beginning larger than the inverse mass, and increases as long as the inverse cascade continues. For simplicity, we approximate the whole infrared region as homogeneous, starting from times when its formation completes in simulations with moderate volumes.
We then extrapolate the dynamics of such homogeneous field by solving the classical field equations:
\beq
\label{meanfield}
\ddot \phi(t) + 3 H \dot \phi(t) + m^2 \phi(t)+\frac{\Lambda^4}{f} \sin\left({ \frac{\phi(t)}{f} }\right) + 2r\frac{\Lambda^4}{f} \sin\left({ 2\frac{\phi(t)}{f} }\right)=0.
\eeq

At high momenta a direct cascade develops after the collapse. As we explain in the following subsections, deviations from the scaling behavior appear when the mass near the bottom of the potential becomes important. At this point the cascade slows down and eventually freezes. To study this effect we evolve the characteristic momentum mode $\bar{\mathrm{p}}(a)$ and the occupation number at that momentum $\bar{n}(a)$ according to a simplified Boltzmann equation, which we derive next.

\subsection{Self-similar evolution in the ultrarelativistic regime} 

As it was mentioned in the main text, there is an extended intermediate stage of the dynamics after fragmentation, during which the potential can be well approximated as massless (conformal) quartic. An important feature of this approximate conformal symmetry is that the evolution equations in the radiation-dominated FRW universe, expressed in terms of comoving coordinates and conformal field and time variables, are the same as those in a Minkowski spacetime. 

The self-similar behavior of the direct cascade at high momenta, where $\omega_{\mathrm{p}} \approx \mathrm{p}$ is satisfied, for the conformal $\varphi^4$ theory has been studied in~\cite{Micha:2004bv}. The scaling exponents of the self-similar evolution,
$$n(\tau,\mathrm{p})= \Bigl( \frac{\tau}{\tau_S} \Bigr)^{\alpha}n_S\Bigl(  \Bigl( \frac{\tau}{\tau_S} \Bigr)^\beta \mathrm{p} \Bigr),$$ are found to be $\beta \approx -1/5$ and $\alpha \approx 4\beta$, where $n_S$ denotes the spectrum at time $\tau_S$. These values can be obtained by solving the Boltzmann equation for a collision term of three-body scattering processes (see also Eq.~(\ref{kineq})). 

Due to the energy cascade the characteristic comoving momentum, which we denote by $\bar{\mathrm{p}}$, grows as $$\bar{\mathrm{p}}(\tau) \propto \tau^{-\beta},$$ with time ($\tau$ is the conformal time). This follows from the form of the self-similar evolution, noting that the characteristic momentum maximizes the energy per mode, $$\mathrm{p}^3n(\tau,\mathrm{p}) = \Bigl( \frac{\tau}{\tau_S} \Bigr)^{\alpha- 3\beta} \Biggl[ \Biggl( \Bigl(\frac{\tau}{\tau_S}\Bigr)^{\beta} \mathrm{p} \Biggr)^3 \,  n_S\Biggl( \Bigl(\frac{\tau}{\tau_S}\Bigr)^{\beta} \mathrm{p} \Biggr)  \Biggr]\: \: \: \: \: \:  \: \: \: \: \: \:\rightarrow \: \: \: \: \: \: \: \: \: \:  \Bigl( \frac{\tau}{\tau_S} \Bigr)^{\beta} \bar {\mathrm{p}}(\tau) = \rm{const}. $$Also the occupation number at that characteristic momentum $\bar{n}$ decreases with time, $$ \bar{n}(\tau) = n(\tau, \bar{\mathrm{p}}(\tau)) =  \Bigl( \frac{\tau}{\tau_S} \Bigr)^{\alpha} \Bigl[  n_S\Bigl(  \Bigl( \frac{\tau}{\tau_S} \Bigr)^{\beta} \bar{\mathrm{p}}(\tau) \Bigr)  \Bigr] \propto \tau^{\alpha}.$$
It is important to keep in mind that the occupation numbers themselves do not decrease due to expansion, since they are defined per comoving volume. This decrease is solely due to the spread of energy to higher momenta.

\subsection{Deviations from the self-similar evolution} 

The self-similar evolution of the energy cascade takes place in the regime of high occupations, $1 \ll \bar{n} \ll \lambda^{-1}$, and  terminates when  $\bar{n}$ becomes of order one. At this stage genuine quantum effects become important and the system is expected to thermalize. This regime of the dynamics is not captured by the classical-statistical approximation, which is instead valid when the spectrum has large occupations at low momenta.

In order to estimate the time when the characteristic occupation numbers would become of order unity, we note that from the above scaling relations it follows that $\bar{n}(\bar{\rm p}) \propto \bar{\mathrm{p}}^{-\alpha/\beta} \propto \bar{\mathrm{p}}^{-4}$. Estimating the proportionality constant from the left panel in Fig.~\ref{spec_rep} we arrive at 
\beq
\bar{n} \sim \frac{1}{\lambda}\Bigl( \frac{30}{\bar{\eta}} \Bigr)^4 = \frac{1}{\lambda}\Bigl( \frac{30 m_a a_{\rm osc}}{\bar{\rm p}} \Bigr)^4 \sim  \frac{1}{\lambda} \Bigl(\frac{a}{15 a_{\rm osc}} \Bigr)^{4\beta},
\eeq
where we also used an estimation $\bar{\rm p}(a) \sim (30 m_a a_{\rm osc}) (a/(15 a_{\rm osc}))^{-\beta}$ from that figure.

On the other hand, the scaling behavior described above is valid only in the ultrarelativistic regime. The small mass near the bottom of the potential eventually leads to deviations from such behavior. These deviations are expected to become important when
\beq
\label{cond_nonrel}
\bar{\rm p} \sim m_{\rm final} a,
\eeq
and, using the above estimate for the characteristic momentum $\bar{\rm p}$, one arrives at
\beq
30 \Bigl(\frac{a}{15a_{\rm osc}}\Bigr)^{-\beta} = r \Bigl(\frac{a}{a_{\rm osc}}\Bigr).
\eeq
Inserting the numerical values for the exponents into the above expressions we obtain parametric estimates for the scale factor for thermalization as well as for the onset of nonrelativistic dynamics,
\beq
\frac{a}{a_{\rm osc} } \sim \frac{15}{\lambda}\: \:  \text{ for thermalization, } \: \: \: \: \: \:  \: \: \frac{a}{a_{\rm osc} } \sim \Bigl(\frac{30}{r}\Bigr) ^{\frac{5}{4}} \: \:  \text{ for the onset of the nonrelativistic regime.}
\eeq 

The cascade slows down soon after the onset of the nonrelativistic behavior, with the field being still far from equilibrium. The reason is that the interaction rates decrease in this regime and become smaller compared to the Hubble expansion rate.

For all values of $\lambda$ and $r$ that are relevant for our ALP models, thermalization happens at much larger scale factors then the transition to the nonrelativistic regime of the dynamics. This implies that the ALPs do not achieve thermal equilibrium when they become nonrelativistic and that the description of this transition can be based on the classical-statistical approximation.

\subsection{Kinetic description and freeze-out}
\label{ssec:kinetic}

In this subsection we demonstrate how the onset of nonrelativistic dynamics, which happens when (\ref{cond_nonrel}) is satisfied, leads to the freeze-out of the cascade. To do this we first write down the kinetic equation describing the cascade, without assuming ultrarelativistic dynamics. We then derive from it a simplified evolution equation for the occupation numbers at the characteristic momentum. We then verify its power-law behavior in the ultrarelativistic regime as well its slow down when the nonrelativistic corrections set in.

The kinetic equation that describes the dynamics of the direct cascade has been derived in~\cite{Micha:2004bv}. It can be written as 
$$ \partial_{\tau} n_{\mathrm{p}} = n'_{\mathrm{p}} H_0 = I_{(3)}+I_{(4)}+... , $$
where $I_{(l)}$ is the collision term corresponding to $l$-body scatterings. As it is shown in~\cite{Micha:2004bv}, the dynamics is dominated by three-body scatterings that arise from the quartic vertex in the presence of a condensate. The corresponding collision term in the regime of large occupation numbers, $n_{\mathbf{p}} \gg 1$, has the form~\cite{Micha:2004bv}
\begin{eqnarray}
\nonumber
\label{kineq}
I_{(3)} = \int \frac{\lambda^2 \langle \phi_c^2 \rangle }{2}   \frac{ d^3\mathrm{k} d^3\mathrm{q}\delta^{(3)}(\mathbf{p-k-q})}{(2\pi)^2  2\omega_{\mathbf{p}} 2 \omega_{\mathbf{k}} 2 \omega_{\mathbf{q}}} \delta(\omega_{\mathbf{0}} + \omega_{\mathbf{p}}- \omega_{\mathbf{k}} - \omega_{\mathbf{q}}) \Bigl[  n_{\mathbf{k}}  n_{\mathbf{q}} -  n_{\mathbf{p}}( n_{\mathbf{k}}+ n_{\mathbf{q}})\Bigr]\\
-2 \int \frac{\lambda^2 \langle \phi_c^2 \rangle}{2} \frac{ d^3\mathrm{k} d^3\mathrm{q} \delta^{(3)}(\mathbf{p+k-q})}{(2\pi)^2 2\omega_{\mathbf{p}} 2\omega_{\mathbf{k}} 2 \omega_{\mathbf{q}}} \delta(\omega_{\mathbf{p}} + \omega_{\mathbf{k}}- \omega_{\mathbf{0}} - \omega_{\mathbf{q}}) \Bigl[  n_{\mathbf{p}}  n_{\mathbf{k}} -  n_{\mathbf{q}}( n_{\mathbf{p}}+ n_{\mathbf{k}})\Bigr],
\end{eqnarray}
where $\omega^2_{\mathbf{p}} = \mathrm{p}^2 + M^2$ is the particle energy and $\langle \phi_c^2 \rangle (\tau)= F_c(\tau, \tau, \mathbf{p}=0)$ describes the condensate in terms of conformal variables.

One can see already from the (\ref{kineq}) why the onset of nonrelativistic regime leads to the slow-down of the dynamics. The particle energies, present in the denominator of the collision integral, grow as $$\omega_{\mathrm{p}} = \sqrt{\mathrm{p}^2 + M^2} \approx M \approx m_{\rm final} a$$ when they become dominated by the rest mass. This leads to the suppression in the right-hand side of (\ref{kineq}).\footnote{In the language of the physical time and the occupation numbers per physical volume the kinetic equation has the form
	$$ \partial_t n_{\rm p} + 3 H n_{\rm p} = C[n]  ,$$
	and the freeze-out happens as a result of the Hubble friction term dominating over the collision term. } Also the $\langle \phi_c^2 \rangle$ term is approximately in the quartic-dominated regime and decreases in the mass-dominated regime.

We now derive the simplified evolution equation for the characteristic modes. Assuming that the shape of the distribution function does not change much throughout the entire dynamics and that the collision term is dominated by momenta close to the characteristic momentum, we can replace the occupation numbers in the squared brackets of (\ref{kineq}) by the characteristic occupation $\bar{n}(\tau)$ and the energies in the denominator by the characteristic energy $\bar{\omega}(\tau) = \omega(\tau, \bar{\mathrm{p}})$. The momentum integrals over the delta-functions give
$$  \int  d^3\mathrm{k} d^3\mathrm{q}\delta^{(3)}(\mathbf{p-k-q}) \delta(\omega_{\mathbf{0}} + \omega_{\mathbf{p}}- \omega_{\mathbf{k}} - \omega_{\mathbf{q}})   =   \int  d^3\mathrm{k} \delta(\omega_{\mathbf{0}} + \omega_{\mathbf{p}}- \omega_{\mathbf{k}} - \omega_{\mathbf{p-k}})  $$ $$ \propto \int  d\mathrm{k} \mathrm{k}^2 \delta(\omega_{\mathbf{0}} + \omega_{\mathbf{p}}- \omega_{\mathbf{k}} - \omega_{\mathbf{p-k}}) \sim  \bar{\mathrm{p}}^2(\tau) \Bigl[ \frac{\bar{\mathrm{p}}(\tau)}{\bar{\mathrm{\omega}}(\tau)} \Bigr]^{-1}. $$
Combining everything leads to the following form of the simplified kinetic equation for the characteristic momentum mode,
\beq
\label{simple_kineq}
\bar{n}' \approx C_{(3)} \frac{  \langle \phi^2_c \rangle   \bar{n} ^2 \bar{\mathrm{p}} }{\bar{\omega}^2},
\eeq
where $C_{(3)}$ is an unknown constant, which encodes the details of the kinetic equation, but should not change much with time. Instead of calculating it we simply find the constant by matching the extrapolation with the lattice simulation.

Of course the derivation of the above equation is not rigorous. Nevertheless (\ref{simple_kineq}) provides an insight into the dynamics. Importantly, the $\bar{\mathrm{p}} \propto \tau^{1/5}$ solution is recovered in the ultrarelativistic regime, $\bar{\omega} \approx \bar{\mathrm{p}}$, under the assumption that $ \langle \phi^2_c \rangle $ stays approximately constant. Moreover, it also describes the slow-down of the cascade once the nonrelativistic behavior sets in, as illustrated in Fig.~\ref{extrapol_1}.

\subsection{Energy conservation for the direct cascade} 

In a static spacetime the direct cascade transports conserved energy. However in an expanding spacetime energy is not conserved. In the language of conformal variables, this is due to the time-dependence of the Hamiltonian, induced by the scale factor. The ultrarelativistic regime in $\varphi^4$ theory is a special case, since the time-dependence drops and a conserved quantity $\rho_c = a^4 \rho$ emerges. 

To be able to extend the discussion about the energy cascades beyond the ultrarelativistic regime we note that in an expanding universe the energy dilutes according to 
\beq
\label{energy_con_exp}
\frac{\rho'}{\rho} = \frac{-3-3w(a)}{a},
\eeq
where the primes denote derivatives with respect to the scale factor. This equation will be used as the generalization of the energy conservation condition in a static spacetime and will allow us to relate the exponents $\alpha$ and $\beta$.

The parameter $w$ can be determined from the energy and the pressure of the field, given by
$$
\rho \approx  \frac{1}{a^4} \int_{\mathbf{p}} \omega(t,\mathrm{p}) n(t,\mathrm{p}), \: \: \: \: \: p \approx  \frac{1}{a^4} \int_{\mathbf{p}} \frac{\mathrm{p}^2}{3\omega(t,\mathrm{p})} n(t,\mathrm{p}).
$$
The energy during the direct cascade is centered around the characteristic momentum, which allows to express the energy and the pressure in terms of $\bar{\mathrm{p}}(a)$ and $\bar{n}(a)$, as well as the corresponding particle energy $\bar{\omega} = \omega(\bar{\mathrm{p}})$, according to 
\beq
\label{eq:pressure_energy}
\rho \propto  \frac{1}{a^4} \bar{\mathrm{p}}^3 \bar{n} \bar{\omega} , \: \: \: \: \: p \propto  \frac{1}{a^4} \bar{\mathrm{p}}^3 \bar{n} \frac{\mathrm{p}^2}{3\bar{\omega}}.
\eeq
The corresponding equation of state parameter $w$ can be estimated from (\ref{eq:pressure_energy}) as
\beq
w = \frac{p}{\rho} = \frac{\bar{\mathrm{p}}^2}{3 \bar{\omega}^2}.
\eeq
Now, from the first equation in (\ref{eq:pressure_energy}) it follows that
$$
\frac{\rho'}{\rho}  = 3 \frac{\bar{\mathrm{p}}'}{\bar{\mathrm{p}}} + \frac{\bar{n}'}{\bar{n}} +\frac{\bar{\omega}'}{\bar{\omega}} - \frac{4}{a}.
$$
Therefore, using (\ref{energy_con_exp}), the energy conservation for the direct cascade can be expressed as 
\beq
\label{energy_conservation}
3 \frac{\bar{\mathrm{p}}'}{\bar{\mathrm{p}}} + \frac{\bar{n}'}{\bar{n}} +\frac{\bar{\omega}'}{\bar{\omega}}  = \frac{1-3w}{a}.
\eeq

Let us consider some particular cases. In the ultrarelativistic regime $\bar{\omega} \approx \bar{\mathrm{p}}$ and $w \approx 1/3$, which leads to the standard energy conservation in $\varphi^4$ theory,$$4\frac{p'}{p} + \frac{\bar{n}'}{\bar{n}}=0.$$ In contrast, in the nonrelativistic regime, where $\bar{\omega} \approx M a$ and $w\approx 0$. As a result, for a constant mass $M$, the third term is approximately $1/a$ and the conservation reads $$ 3 \frac{p'}{p} + \frac{\bar{n}'}{\bar{n}} =0. $$

\bigskip

To summarize the appendix, we solve a set of coupled equations for the homogeneous component (\ref{meanfield}), for the characteristic momentum scale (\ref{simple_kineq}) and the one reflecting energy conservation (\ref{energy_conservation}). These equations are supplemented by the approximate equation for the particle energy at the characteristic momentum $\bar{\omega}(a)$,
\beq
\label{simple_omega}
\bar{\omega}^2 = \bar{\rm p}^2 +m_{\rm final}^2 a^2 +\frac{\lambda \langle \phi_c^2 \rangle}{2}.
\eeq
We initialize the variables $\phi_c$, $\bar{n}$ and $\bar{\rm p}$ as well as the prefactor $C_{(3)}$ using the data from lattice simulations. The results from this analysis are summarized in Sec.~\ref{extrapol} and, in particular, in figures~\ref{extrapol_1} and~\ref{extrapol_2}.

\bibliographystyle{utphys}

\bibliography{references}

\end{document}